\documentclass[11pt]{JHEP3}

\usepackage{amsmath}
\usepackage{amsfonts}
\usepackage{amssymb}
\usepackage{latexsym}
\usepackage{slashed}
\usepackage{mathrsfs}
\usepackage{cancel}
\usepackage{graphicx}

\numberwithin{equation}{section}
\numberwithin{table}{section}

\newcommand{\be}{\begin{equation}}
\newcommand{\ee}{\end{equation}}
\newcommand{\bes}{\begin{equation*}}
\newcommand{\ees}{\end{equation*}}
\newcommand{\bea}{\begin{eqnarray}}
\newcommand{\eea}{\end{eqnarray}}
\newcommand{\beas}{\begin{eqnarray*}}
\newcommand{\eeas}{\end{eqnarray*}}
\newcommand{\bb}{\mathbb}

\newcommand{\p}{\partial}

\renewcommand{\tfrac}[2]{{\textstyle\frac{#1}{#2}}}

\def\Z{\mathbb{Z}}

\newcommand{\Qb}{{\overline{Q}}}

\newcommand{\Gam}{\Gamma}

\def\cN{\mathcal{N}}

\newcommand{\dalf}{{\dot{\alpha}}}

\newcommand{\alf}{\alpha}
\newcommand{\bet}{\beta}

\newcommand{\gam}{\gamma}
\newcommand{\lam}{\lambda}
\newcommand{\Lam}{\Lambda}

\newcommand{\sig}{\sigma}
\newcommand{\sigb}{\overline{\sigma}}
\newcommand{\chib}{\bar{\chi}}

\def\Im{\mathrm{Im}}

\def\R{\mathbb{R}}
\def\ub{\bar{u}}
\def\tb{\bar{t}}

\newcommand{\ib}{{\bar{\imath}}}
\newcommand{\jb}{{\bar{\jmath}}}
\newcommand{\kb}{{\bar{k}}}
\newcommand{\lb}{{\bar{l}}}
\newcommand{\mb}{{\overline{m}}}

\def\ab{\bar{a}}

\def\db{\bar{d}}
\def\bb{\bar{b}}
\newcommand{\Fb}{\overline{F}}
\def\zb{\bar{z}}

\def\P{\mathbb{P}}
\def\C{\mathbb{C}}

\def\hK{\hat{K}}
\def\ads{AdS$_4$}

\def\Wb{\overline{W}}
\def\cL{\mathcal{L}}

\def\cO{\mathcal{O}}

\def\yb{\bar{y}}
\def\hX{\hat{X}}

\def\wb{\bar{w}}

\def\hK{\hat{K}}
\def\xib{\bar{\xi}}

\def\smap{\varphi}    
\def\sc{\smap}           
\def\csc{\smap}         
\def\bcsc{\bar\csc}    
\def\Sfl{P}                 
\def\bSfl{\overline{\Sfl}}
\def\Ssugra{P}          
\def\bSsugra{\overline{\Ssugra}}
\def\Sads{W}            
\def\bSads{\overline{\Sads}}
\def\smodel{sigma model}

\title{$\cN=1$ Sigma Models in AdS$_4$}

\preprint{MIT-CTP-4246, SU-ITP-11/21, NSF-KITP-11-053}

\author{Allan Adams$^1$, Hans Jockers$^{2,3}$, Vijay Kumar$^3$, Joshua M. Lapan$^3$\\
$^1$
Center for Theoretical Physics\\
Massachusetts Institute of Technology\\
Cambridge, MA 02139, USA\\
\\
$^2$
Department of Physics\\ 
Stanford University\\ 
Stanford, CA 94305-4060, USA\\
\\
$^3$
Kavli Institute for Theoretical Physics\\
University of California\\
Santa Barbara, CA 93106, USA \\
\\
{\tt awa} {\rm at} {\tt mit.edu}, {\tt jockers} {\rm at} {\tt kitp.ucsb.edu}, {\tt vijayk} {\rm at} {\tt kitp.ucsb.edu}, {\tt lapan} {\rm at} {\tt kitp.ucsb.edu}
}

\abstract{
We study sigma models in AdS$_4$ with global $\cN=1$ supersymmetry and find that they differ significantly from their flat-space cousins --- the target space is constrained to be a K\"ahler manifold with an exact K\"ahler form,
the superpotential transforms under K\"ahler transformations,
the space of supersymmetric vacua is generically a set of isolated points even when the superpotential vanishes, and the $R$-symmetry is classically broken by the cosmological constant.
Remarkably, the exactness of the K\"ahler class is also required for the sigma model to arise as a decoupling limit of $\cN=1$ supergravity, and ensures the vanishing of gravitational anomalies.
As applications of these results, we argue that fields with AdS$_{4}$ scale masses are ubiquitous in, for example, type IIB $\cN=1$ AdS$_{4}$ vacua stabilized near large volume; we also present a schematic argument that the Affleck-Dine-Seiberg runaway of $N_f<N_c$ SQCD can be regulated by considering the theory in AdS$_{4}$.
}

\begin{document}














\newpage

\parskip 0.1in
\parindent 0.3in

\section{Introduction and Conclusions}

\def\ie{{i.e.}}

The considerations in this paper are driven by two general observations.
First is the curious fact that large radius compactifications to four-dimensional anti-de Sitter space with $\cN=1$ supersymmetry very often come with anomalously light moduli whose masses scale with the AdS$_{4}$ curvature (see \cite{Douglas-Kachru-review, Blumenhagen:2006ci, Denef-les-houches} for reviews of moduli stabilization scenarios and AdS$_{4}$ vacua in string theory).   
This is not a deep obstruction to phenomenological model building, as one can certainly give all moduli parametrically large masses by for example breaking supersymmetry or stepping away form a perturbative large-radius limit, which successful models such as KKLT \cite{KKLT} of course do.  Nevertheless, it is instructive to take this issue seriously as it will reveal features of rigid sigma models in AdS$_4$, and their coupling to supergravity, that depart from naive flat-space intuition.

Second, 
it has recently been argued
that consistent flat-space decoupling limits of supergravity lead to very special rigid supersymmetric theories \cite{KS}.
More precisely, it has been argued that the target space of an $\cN=1$ sigma model in flat space must have an exact K\"ahler form in order to couple it to linearized supergravity; when applied to typical classes of string compactifications, this implies the inevitable existence of massless moduli in any smooth decoupling limit to flat space that preserves supersymmetry.  It would be interesting to understand how and when such arguments apply in supersymmetric AdS$_4$ compactifications.

In this paper, we will argue that both of these properties --- the ubiquity of light moduli in $\cN=1$ AdS$_4$ compactifications and the constraints on the topology and geometry of sigma models arising in the rigid limit of supergravity theories --- follow from basic properties of 
supersymmetry in AdS$_4$.  
More precisely, we will study $\cN=1$ AdS$_4$ sigma models with K\"ahler target spaces $X$ both 
as theories with global AdS$_4$ supersymmetry 
and as decoupling limits of consistent supergravity theories.   As we shall explain, both of the above 
observations
follow from simple, but surprisingly constraining, consistency conditions and kinematic properties of rigid AdS$_4$ sigma models.

Let us briefly summarize these constraints and properties.  
A supersymmetric sigma model in \ads\ (with cosmological constant $-3\lam^2$) is specified by a K\"ahler target space $X$ with K\"ahler potential $K(\csc, \bcsc)$, and by a holomorphic superpotential $W(\csc)$ governing relevant interactions.
 A key property of these rigid theories in AdS$_4$ is that, as in supergravity, the superpotential is not an invariant object but rather mixes with the K\"ahler potential under K\"ahler transformations,
$$
  K(\csc,\bcsc) \rightarrow K(\csc,\bcsc) + f(\csc) + \bar f(\bcsc) \ , \qquad
  \Sads(\csc) \rightarrow  \Sads(\csc) - \lam \, f(\csc) \,.
$$
The usual flat-space formula for the scalar potential, $g^{i\jb} W_i \Wb_\jb$, is not invariant under K\"ahler transformations, and must therefore be modified in AdS$_4$. Indeed, we find the form
$$
V(\csc, \bcsc) =  g^{i\jb}(\Sads_i + \lam K_i )(\bSads_\jb+\lam K_\jb) - 3\lam \bSads - 3\lam\Sads - 3 \lam^2 K\ \, .
$$
The K\"ahler invariance of the action leads to a host of constraints on the possible form of the theory.  
In particular, supersymmetry further requires that the target $X$ have a {\em trivial K\"ahler class}, $\left[\omega\right]=0$.  If we wish to build a sigma model starting with a manifold containing compact holomorphic cycles, we must fiber additional scalars over the geometry to trivialize all these cycles; otherwise, the theory itself will spontaneously decompactify them.  If we started with a Hodge manifold then one such scalar would suffice, but generally we are forced to introduce $h^{1,1}(X)$ independent scalars.  

Interestingly, the consistency conditions we find for the target space of a rigid $\cN=1$ sigma model in AdS$_4$ are (a) identical to those required for a rigid {\em flat-space} sigma model to arise as a decoupling limit of supergravity \cite{KS}, and (b) imply the vanishing of all (mixed) gravitational anomalies upon coupling the rigid sigma model to $\cN=1$ supergravity.  Thus, working in a non-trivial classical background makes this quantum constraint classically manifest, as in \cite{AdamsNima}.  It is pleasing to see these constraints go over smoothly as $\lambda\to0$.

A second surprising fact follows from the mixing of the K\"ahler potential and superpotential in rigid AdS$_{4}$: the right hand side of the supersymmetry variations are proportional not simply to $\Sads_i$, as in flat space, but to $\Sads_i + \lam K_i $.  Thus, even when the superpotential is zero, the vanishing of the fermion variations impose $n$ equations on the $n$ sigma model coordinates, implying that the supersymmetric vacua of this sigma model are generically a set of isolated points on $X$!\footnote{For non-generic models, there can be flat directions.}  This differs sharply from familiar intuition from flat-space sigma models, where the moduli space for a vanishing superpotential is the full manifold, $X$.  However, these two results pair naturally: away from these isolated points, the scalar potential is non-zero but scales as $\lambda$ --- the supersymmetric points being gentle attractors --- implying that we recover the expected moduli space in the flat-space limit.\footnote{Note that we can, of course, turn on additional superpotential terms to shift the masses of the light fields and shift around the supersymmetric points; we cannot, however, make the vanishing of the fermion variations a holomorphic condition, unlike in flat space, unless the K\"ahler potential is special K\"ahler (or simply trivial).  This raises an interesting question about $\cN=2$ sigma models in AdS$_4$, but that is beyond the scope of this discussion.}  It is 
useful to think of AdS$_{4}$ as a homogenous box inside of which massless modes are gapped: even if the target space is non-compact, the zero mode on the target feels a harmonic potential due to the constant negative curvature and is therefore massive.

The results above can be used to build a cartoon 
argument for the ubiquity of light moduli in large-radius compactifications respecting an unbroken ${\cal N}=1$ supersymmetry.  (Such ${\cal N}=1$ compactifications play important roles in a variety of moduli stablization scenarios, including for example the KKLT scenario \cite{KKLT}, where the ensuing light moduli can subsequently be lifted by the supersymmetry-breaking uplifting stage.) 
Consider a large-radius flux compactification on some Calabi-Yau (or a decorated version thereof) to AdS$_{4}$ with suitably small cosmological constant, $\lambda$.  By ``large radius'' we mean a manifold that is large compared to the four-dimensional Planck scale, allowing for a consistent perturbative expansion in $1/M_{pl}$ around the rigid limit; by ``suitably small'' we mean $\lambda\ll M_{pl}$.  

At leading order in this perturbative expansion (\ie\ \!\!, in the decoupling limit), the results above tell us that there must be light fields in the theory with masses of order  $\lambda$ as long as supersymmetry is unbroken. Now, if perturbation theory in $\lambda/M_{pl}$ is valid, 
leading corrections from supergravity 
will shift the masses of these light moduli by $\cO\Big({\lambda \over M_{pl}}\Big)$ --- \ie\ \!\!, not by very much.  For these pesky moduli to be lifted above the AdS$_{4}$ scale, perturbation theory around large radius must {\em not} be reliable --- we would need corrections that are nonperturbative in $1/M_{pl}$ --- in which case we should not 
overcommit ourselves to 
perturbation theory around large radius with unbroken $\cN=1$ supersymmetry.  This is, of course, what successful models of stabilization already do, 
for example by breaking supersymmetry and uplifting.
In large volume compactifications of type IIB with $\cN=1$ AdS$_4$ vacua, we will show that the moduli are light with masses necessarily proportional to $\lam$.

The fact that sigma models in AdS$_4$ have moduli spaces composed of isolated points suggests that the study of $\cN=1$ gauge theories may also considerably simplify in AdS$_4$.  We shall present a schematic argument to this effect by studying the example of $SU(N_{c})$ SQCD with $N_{f}<N_{c}$ fundamental quarks in AdS$_4$.  In flat space, this theory famously suffers from an Affleck-Dine-Seiberg (ADS) runaway \cite{Affleck-Dine-Seiberg} in which various mesons run off to infinity in field space.  In AdS$_4$, however, the zero mode of a sigma model is generically lifted, suggesting that the ADS runaway may be lifted in AdS.  We will marshall evidence for this picture, arguing that  the meson field should be stabilized at a finite vev controlled by the ratio of the confinement scale $\Lambda_{c}$ to the cosmological constant, $\langle m\rangle \sim \left({\Lambda_{c}\over\lambda}\right)^{\frac{1}{2}}$.  A detailed study of $\cN=1$ gauge theories in AdS$_4$ is beyond the scope of this paper but is, clearly, of considerable interest.\footnote{See \cite{Tong} for a discussion of Seiberg-Witten theory in AdS. We thank D. Tong for interesting and valuable discussions on these and related issues.}

We thus see that simple consistency conditions for $\cN=1$ sigma models in AdS$_4$ lead to the constraints for weakly coupling to supergravity in the flat space limit \cite{KS}, to the ubiquity of light moduli in large-radius $\cN=1$ compactifications to AdS$_{4}$, and to a surprising set of features of rigid $\cN=1$ sigma models and gauge theories in AdS$_{4}$.  The remainder of the paper is devoted to deriving and explicating the above results, approaching them from two directions.  First, in Section 2, we study the structure of rigid $\cN=1$ sigma models in AdS$_4$, deriving many of their properties directly and exploring a set of illustrative examples.  
We also show how the AdS$_{4}$ supersymmetric Lagrangian can be derived from supergravity through a decoupling limit.
In Section 3, we discuss the constraints on sigma models, particularly the triviality of the K\"ahler class, in the context of work by Bagger \& Witten \cite{Witten:1982hu} and Komargodksi \& Seiberg \cite{KS} on the decoupling limits of supergravity theories. We also prove that the triviality of the K\"ahler class implies that there are no mixed gravitational anomalies when the sigma model is coupled to supergravity. The $[\omega]=0$ constraint can, therefore, be derived by looking at mixed gravitational anomalies around a flat background, or through purely classical considerations of sigma models in an AdS$_{4}$ background. We 
suggest
that this coincidence is not an accident but a consequence of a more general ``Background Principle'', which we briefly discuss. 
In Section 4, we present a simple argument for the existence of AdS$_{4}$ scale moduli in string compactifications at large volume that preserve supersymmetry; this suggests that one must break supersymmetry, or move away from large volume, in order to give large masses to the moduli, 
as is of course done in all phenomenologically successful models of moduli stabilization such as KKLT.
We end in Section 5 with a discussion of how the low-energy behavior of supersymmetric gauge theories, specifically the Affleck-Dine-Seiberg runaway, may be altered by AdS.

\section{Supersymmetric Lagrangians for Chiral Multiplets in AdS$_4$} 
\label{sec:ads4-sigma}

In this section, we construct the most general supersymmetric Lagrangian describing the interactions of chiral multiplets in \ads. 
 We show that the superpotential shifts under K\"ahler transformations and then derive various consequences from this fact: generally, there are no moduli spaces of supersymmetric vacua in AdS$_4$; the K\"ahler class of the target space must be trivial. We also present an alternate derivation of the sigma model Lagrangian as a decoupling limit of supergravity. 
 For earlier work on rigid supersymmetric quantum field theories in AdS$_4$, see \cite{Keck:1974se, Zumino:1977av, Ivanov:1979ft, Ivanov-Sorin, BF, Burges:1985qq, deWit-Herger,  McKeon:2003xb, Gripaios:2008rg, Rattazzi:2009ux, Aharony:2010ay}.

\subsection{The $\cN=1$ Supersymmetric Sigma Model in Minkowski Space}
\label{sec:sigma-review}

The well-known four-dimensional \smodel\ describes the general effective action (usually up to two derivatives) governing the interactions of massless scalar fields. It is given by a map $\smap: M \rightarrow X$ from the spacetime $M$ into a target space $X$, which is also equipped with a (positive-definite) metric $g$. Then the \smodel\ action reads
\be\label{eq:Svarphiinv}
   S_{kin} = f_\pi^2 \int d^4x \sqrt{-\gam}\, \gam^{mn}\,(\varphi^*g)(\partial_m,\partial_n) \ ,
\ee
which in local target-space coordinates, $\sc^I$, corresponding to the (massless) scalar fields of the \smodel, has the familiar form
\be \label{eq:nonsusy-sigma}
  S_{kin} = f_\pi^2\int d^4x\sqrt{-\gam}\, g_{IJ}(\sc) \partial_m \sc^I \partial^m \sc^J \ .
\ee
Here $\gam_{mn}$ is the spacetime metric with respect to the basis of tangent vectors $\partial_m$, and $\varphi^*g$ denotes the pullback of the metric $g=g_{IJ}\ d\sc^Id\sc^J$ of the target space $X$ to the spacetime $M$.
There is a characteristic energy scale, $f_\pi$, that controls the strength of the scalar self-interactions. Due to unitarity constraints on low-energy scattering amplitudes (discussed, for example, in \cite{Donoghue:1992dd}), the effective action has a UV cutoff $\Lam_\sig \lesssim 4\pi f_\pi$.

The mass dimensions of the \smodel\ fields are $[\sc^I]=0, \ [\partial_m] = 1, \ [f_\pi]=1$. The kinetic term \eqref{eq:nonsusy-sigma} (without the normalization factor $f_\pi^2$) has mass dimension two. We can include interaction terms with mass dimension less than or equal to two, which are then relevant or marginal with respect to the scalar kinetic term (e.g., mass terms $(m^2)_{IJ}\varphi^I\varphi^J$). 

To set the stage, and for later reference, we collect some well-known properties of the $\cN=1$ sigma model in Minkowski space $\mathbb{R}^{1,3}$, which describes the interactions of $\cN=1$ chiral multiplets. As before, the kinetic term for the bosonic scalars in the chiral multiplets is given by the action~\eqref{eq:Svarphiinv}, but $\cN=1$ supersymmetry requires the target space $X$ to be a complex K\"ahler manifold with K\"ahler metric $g$ \cite{Wess:1992cp}. In local complex target space coordinates $\csc^i$, which are identified with the complex scalars $\csc^i$ of the $\cN=1$ chiral multiplets, the bosonic kinetic term reads
\be \label{eq:susy-sigma}
S^{bos}_{kin} = f_\pi^2\int d^4x \, g_{i\bar\jmath}(\csc,\bar\csc) \, \partial_m \csc^i \partial^m \bcsc^{\bar\jmath} \ .
\ee
The K\"ahler potential $K(\csc,\bcsc)$ is a real function of the complex scalars $\csc$, and it specifies locally the K\"ahler metric $g$ and the K\"ahler $(1,1)$-form $\omega$ as
\be \label{eq:Kmet}
  g_{i\bar\jmath}(\csc,\bcsc) = \frac{\partial^2}{\partial \csc^i\partial\bcsc^{\bar\jmath}}K(\csc,\bcsc) \ , \qquad
  \omega_{i\bar\jmath}(\csc,\bcsc) = i\,\frac{\partial^2}{\partial\csc^i\partial\bcsc^{\bar\jmath}}K(\csc,\bcsc) \ .
\ee
Note that the K\"ahler metric $g$, the K\"ahler form $\omega$, and consequently the supersymmetric \smodel\ itself, are invariant under K\"ahler transformations
\be \label{eq:Ktrans}
  K(\csc,\bcsc) \rightarrow K(\csc,\bcsc) + f(\csc) + \bar f(\bcsc) \ ,
\ee 
for arbitrary holomorphic functions $f(\csc)$. 

The K\"ahler potential allows us to express the whole supersymmetric \smodel\ action (also including the fermionic terms) in global $\cN=1$ superspace \cite{Wess:1992cp}
\be \label{eq:SuperKin}
  S_{kin} = f_\pi^2 \int d^4x\,d^4\theta\,K(\Phi,\bar\Phi)  \ .
\ee
Here, the arguments of the K\"ahler potential are the chiral superfields $\Phi$ associated to the complex scalars $\csc$ and the integral is taken over the whole $\cN=1$ superspace.

Relevant interactions in the $\cN=1$ supersymmetric \smodel\ (such as mass terms) are encoded in the superpotential $\Sfl$, which is a holomorphic function on the K\"ahler target space $X$. In the action, the superpotential $\Sfl$ yields (locally) the bosonic interaction terms
\be
  S_{int}^{bos} = f_\pi^2 \int d^4x \,g^{i\bar\jmath}(\csc,\bcsc) \Sfl_i(\csc) \bar\Sfl_{\bar\jmath}(\bcsc) \ ,
\ee
with $\Sfl_i \equiv \frac{\partial\Sfl}{\partial\csc^i}$. These bosonic interactions pair with the fermionic terms and are conveniently expressed in terms of the $\cN=1$ superspace superpotential interaction
\be \label{eq:SuperInt}
  S_{int} = f_\pi^2 \int d^4x \, d^2\theta \, \Sfl(\Phi) + {\rm c.c.} \ .
\ee  

The K\"ahler potential $K(\csc,\bcsc)$, defined on a local patch in the target space $X$, need not be extendable to a function that is well-defined over the entire space $X$. Instead, in order to yield a globally well-defined K\"ahler metric $g$, the various local K\"ahler potentials $K$ may differ by K\"ahler transformations \eqref{eq:Ktrans} on overlapping regions of the local patches. In fact, the obstruction to extending a local K\"ahler potential to one defined over all of $X$ (without the use of K\"ahler transformations on overlaps) is measured by the Dolbeault cohomology class in $H^{1,1}(X)$ of the K\"ahler $(1,1)$-form $\omega$. Thus, unless the form $\omega$ is exact --- i.e., $\omega = d\theta$ globally --- the K\"ahler manifold $X$ does not admit a globally defined K\"ahler potential $K$.  

Note that an exact K\"ahler form has strong implications on the K\"ahler target space geometry and topology.  Recall that compact holomorphic submanifolds of K\"ahler manifolds are calibrated by the K\"ahler class $\omega$ --- i.e., the volume of a compact holomorphic submanifold $S$ of complex dimension $n$ is given by (e.g., \cite{Harvey:1982})
\be \label{eq:Kcali}
  {\rm vol}(S) = \frac{1}{n!} \int_S \omega^n \ .
\ee  
This innocent looking property has important consequences for K\"ahler manifolds with an exact K\"ahler form $\omega = d\theta$: exactness of the integrand implies that the integral of $\omega^n$ over any compact submanifold vanishes. For exact K\"ahler forms, the calibration condition~\eqref{eq:Kcali} thus implies that the only compact holomorphic submanifolds are points (since any compact submanifold of dimension greater than zero must have a finite volume), a consequence of which is that {\em the K\"ahler manifold itself must be non-compact}.


\subsection{Chiral Multiplets in \ads}
Consider \ads\ with radius $\lambda^{-1}$, e.g., as a hyperboloid $-x_-^2 -x_0^2 + x_1^2 + x_2^2+x_3^2=-\frac{1}{\lam^2}$ embedded in $\R^{2,3}$.  The associated $\cN=1$ \ads\ superalgebra, denoted by $osp(1,4)$,\footnote{We can, of course, remove $\lambda$ from the supersymmetry algebra by a simple rescaling, $Q_{\alpha}\to\sqrt{\lambda} \, Q_{\alpha}$ and $R_{a}\to\lambda R_{a}$.  We work with the chosen normalization to make the flat space limit manifest.} reads \cite{Keck:1974se,Zumino:1977av,Ivanov-Sorin}
\be \label{eq:Ads4algebra}
\begin{aligned}
\{Q_\alf , \Qb_\dalf\} &= -2\sig^a_{\alf\dalf} R_a \ , \qquad & \{Q_\alf, Q^\bet\} &= 2i\lambda {(\sig^{ab})_\alf}^\bet M_{ab} \ ,  \\
[R_a, Q_\alf] &= -\frac{1}{2} \lambda (\sig_a \Qb)_\alf \ , & [M_{ab}, Q_\alf] &= -i {(\sig_{ab})_\alf}^\beta Q_\bet \ , \\
[R_a,R_b] &= -i\lambda^2 M_{ab} \ , & [M_{ab},R_c]&= i(\eta_{ac} R_b - \eta_{bc} R_a)  \ .
\end{aligned}
\ee
The vector indices $a, b,\ldots,$ and the spinor indices $\alf, \dalf,\ldots,$ refer to the local Lorentz frame, and we use Wess and Bagger \cite{Wess:1992cp} two-component spinor notation throughout. $Q_\alf$ and $\Qb_\dalf$ are the $\cN=1$  fermionic supersymmetry generators, whereas the bosonic generators $R_a$ and $M_{ab}$ 
generate AdS$_{4}$ translations and local Lorentz transformations, respectively.  These operators combine into the generators $(\lam R_a, M_{ab})$, which are the generators of the group $SO(2,3)$ that acts on the \ads\ hyperboloid. Note that in the limit $\lambda\rightarrow 0$, the generators $R_a$ and $M_{ab}$ become the usual translation and rotation operators of four-dimensional Minkowski space and the \ads\ superalgebra \eqref{eq:Ads4algebra} reduces to the familiar four-dimensional $\cN=1$ superalgebra of flat space.

Starting from the superalgebra \eqref{eq:Ads4algebra}, we can derive the representations on the fields following, for instance, \cite{Weinberg:2000cr}. The supersymmetry transformation of a field $\Phi$ is defined as
\begin{equation}
\delta_\xi \Phi (x) = -i[\xi Q + \xib \Qb, \Phi(x)] \ .
\end{equation}
The chiral multiplet of AdS$_4$ supersymmetry is defined as a multiplet whose lowest component is a complex scalar that is annihilated by $\Qb_\dalf$, \ie, $[\Qb_\dalf, \csc] = 0$. By acting on the lowest component with the ``raising operator'' $Q_\alf$, we can derive the other components of the multiplet and their supersymmetry transformations. As in flat space, the chiral multiplet consists of a complex scalar $\csc$, a Weyl fermion $\chi$, and a complex auxiliary field $F$.  The transformation laws are listed below:
\be \label{eq:ads-susy-var} 
\begin{aligned} 
\delta_\xi \csc^i & = \sqrt{2} \, \xi \chi^i \ , \\
\delta_\xi\chi^i & = \sqrt{2}\, F^i \xi  +i \sqrt{2}\, \sig^m \xib \p_m \csc^i \ ,  \\ 
\delta_\xi  F^i & = -\sqrt{2}\, \lambda \xi\chi^i + i\sqrt{2}\, \xib \sigb^m \nabla_m \chi^i  \ .
\end{aligned}
\ee
The algebra closes on these fields only if the supersymmetry parameter $\xi$ satisfies the Killing spinor equation,
\be \label{eq:Killing}
(\nabla_m \xi)^\alf = \frac{i\lam}{2} (\xib \sigb_m)^\alf, \qquad (\nabla_m \xib)_\dalf = \frac{i\lam}{2} (\xi \sig_m)_\dalf \ .
\ee
In the $\lam \rightarrow 0$ limit, when the AdS$_4$ superalgebra reduces to the Poincar\'e superalgebra, the transformations \eqref{eq:ads-susy-var} reduce to the usual supersymmetry transformations of a chiral multiplet in flat space, with the supersymmetry parameter $\xi$ a constant spinor. 

Note that the \ads\ superalgebra \eqref{eq:Ads4algebra} does not enjoy an $R$-symmetry.  The trouble is particularly clear in the Killing spinor equation, (\ref{eq:Killing}), which relates $\xi$ to $\bar{\xi}$ and thus does not allow a chiral rotation of $\xi$.  However, as long as we have a consistent flat space limit, the broken $R$-symmetry can still be useful.  To see this, imagine our rigid \ads\ theory came from a decoupling limit of some $\cN=1$ supergravity (we discuss this in more detail in Section \ref{sec:SUGRAred}).  In this case,  $\lambda$ is the vev of the superpotential, $\lambda = M_{pl}^{-2}\langle P \rangle $.  Since the superpotential carries $R$-charge 2, this vev breaks the $R$-symmetry.  We may thus treat $\lambda$ as an $R$-charge 2 spurion controlling the breaking of $R$-symmetry --- indeed, this charge assignment restores the $R$-covariance of the killing spinor equation (\ref{eq:Killing}) while also forbidding the rescaling discussed above that removes $\lambda$ from the supersymmetry algebra.

Just how far such a spurion analysis can take us is an extremely interesting question.  For example, can we use $\lambda$ to define an invariant holomorphic superpotential even away from a flat space limit?  Can we use the would-be $R$-symmetry to constrain patterns of supersymmetry breaking, extending the arguments of Nelson and Seiberg \cite{Nelson:1993nf}\ on spontaneous supersymmetry breaking in Wess-Zumino models to the \ads\ context?  Do standard arguments, which rely heavily on (possibly anomalous) $R$-symmetries,  extend to $\cN=1$ theories in \ads?  
For now, we simply note these questions and postpone a more detailed discussion to future work.\footnote{A.A. thanks J. Thaler for discussions on this and related topics.}


\subsection{\ads\ Supersymmetric Sigma Model}

Quantum field theories in \ads\ are naturally regulated in the IR by the AdS$_{4}$ scale $\lam$ (see \cite{Callan:1989em} for an excellent discussion).  In the remainder of this section, we focus on $\cN=1$ sigma models in \ads, which also enjoy a UV cutoff, $4\pi f_\pi$, with a hierarchy thus given by 
\be
\lam \ll \Lam_\sig \lesssim 4\pi f_\pi \ .
\ee
To derive the most general couplings of chiral multiplets in \ads, we first write down all the possible terms with mass dimension no greater than two. The relevant assignments of the mass dimensions are given by
\begin{equation}
[\csc^i] = 0\ , \quad [\chi^i] = \frac{1}{2}\ , \quad [F^i] = 1 \ , \quad [\lam] = 1 \ , \quad [\partial_m] = 1\ . \label{eq:sig-scaling}
\end{equation}
Since in the $\lam \rightarrow 0$ limit the AdS$_{4}$ superalgebra becomes the usual Poincar\'e superalgebra, we expect that the Lagrangian for the supersymmetric \smodel\ in \ads\ smoothly goes over to the supersymmetric \smodel\ Lagrangian in flat space. The flat-space Lagrangian (including interaction terms induced from the superpotential $\Sfl(\csc)$) is given by the superspace action \eqref{eq:SuperKin} and \eqref{eq:SuperInt}, which yields the component expression \cite{Wess:1992cp}
\bea
\cL (\lam \rightarrow 0) & = & g_{i\jb} F^i \Fb^\jb  - g_{i\jb} \partial_m \csc^i \partial^m \bcsc^\jb - i g_{i\jb} \chib^\jb  \sigb^m \mathfrak{D}_m\chi^i  - \frac{1}{2} \Sfl_{ij} \chi^i \chi^j - \frac{1}{2} \bar\Sfl_{\ib\jb} \chib^\ib \chib^\jb \nonumber \\
& & - F^i \left( \frac{1}{2} g_{i\jb,\kb} \chib^\jb \chib^\kb  - \Sfl_i\right) - \Fb^\ib \left( \frac{1}{2} g_{j\ib, k} \chi^j \chi^k - \bar\Sfl_\ib \right) + \frac{1}{4} g_{i\jb, k\lb} \chi^i\chi^k\chib^\jb\chib^\lb \ . \nonumber
\eea
Here $\mathfrak{D}_m \chi^i \equiv \partial_m \chi^i + \partial_m \csc^k \Gamma^i_{jk} \chi^j$ is the K\"ahler covariant derivative with respect to the K\"ahler target space connection and, as in \eqref{eq:Kmet}, the K\"ahler metric can locally be derived from a K\"ahler potential $K$. 

According to (\ref{eq:sig-scaling}), all terms in the above Lagrangian have scaling dimension less than or equal to two. In the \ads\ background, the dimensionful parameter $\lam$ allows for other relevant terms.  Requiring that the Lagrangian has a smooth $\lam \rightarrow 0$ limit, we arrive at the following ansatz
\bea \label{eq:Lag-ansatz}
\cL( \lam) & = & g_{i\jb} F^i \Fb^\jb  - g_{i\jb} \partial_m \csc^i \partial^m \bcsc^\jb - i g_{i\jb} \chib^\jb  \sigb^m \mathfrak{D}_m \chi^i  - \frac{1}{2} \Sfl_{ij} \chi^i \chi^j - \frac{1}{2} \bSfl_{\ib\jb} \chib^\ib \chib^\jb \nonumber \\
& & - F^i \left( \frac{1}{2} g_{i\jb,\kb} \chib^\jb \chib^\kb  - \Sfl_i\right) - \Fb^\ib \left( \frac{1}{2} g_{j\ib, k} \chi^j \chi^k - \bSfl_\ib \right) + \frac{1}{4} g_{i\jb, k\lb} \chi^i\chi^k\chib^\jb\chib^\lb \nonumber \\
& & + \lam (r + F^i t_i + \Fb^\ib \tb_\ib + u_{ij} \chi^i \chi^j + \ub_{\ib\jb} \chib^\ib \chib^\jb) \ .
\eea
The additional terms on the last line are all proportional to $\lam$. 
The new parameters we have introduced, $r,  t_i, u_{ij}$, are all  functions of $\csc^i$ and $\bcsc^\jb$. 
The parameter $r$ is real with mass dimension one, while $t_i$ and $u_{ij}$ are complex with mass dimension zero. 
Also note that in the $\chi$ kinetic term, the covariant derivative $\mathfrak{D}_m$ now includes \ads\ spin-connection in addition to the target-space connection.

We now demand that the above Lagrangian be invariant under the \ads\ supersymmetry variations \eqref{eq:ads-susy-var}. Acting with these supersymmetry variations on the Lagrangian \eqref{eq:Lag-ansatz}, we find, 
\be \label{eq:LagVar}
\begin{aligned}
\frac{1}{\sqrt{2}} \delta_\xi \cL = \lam 
\Big[ & \xi\chi^i\left( r_i - 3 \Sfl_i -3 \lam t_i \right) + (\xi\chi^i)(\chib^\jb\chib^\kb)\left(\tfrac12 g_{i\jb,\kb} + \ub_{\jb\kb,i}\right)
+ (\xi\chi^i)(\chi^j\chi^k) u_{jk,i} \\
& + \xi\chi^i F^j\left(t_{j,i}+2 u_{ij}\right) + \xi\chi^i\Fb^\jb \left( -g_{i\jb} + \tb_{\jb,i} \right) \\
&+ i \xi\sigma^m\chib^\ib \left(\nabla_m\csc^j(g_{j\ib}-t_{j,\ib})+\nabla_m\bcsc^\jb(\tb_{\jb\ib}+2 \ub_{\ib\jb}) \right) \Big] + 
{\rm c.c.} \ .
\end{aligned}
\ee
In deriving this variation we have integrated by parts and used the Killing spinor equation \eqref{eq:Killing}.\footnote{Throughout this paper, we will be cavalier about the boundary conditions satisfied by our fields.  At the present step, we do not expect any obstructions to finding boundary conditions that are compatible with the supersymmetry variations described.  In other steps, however, we must be more careful since the necessity of choosing certain boundary conditions can be surprisingly restrictive; for example, they can be shown to forbid the existence of charged chiral fermions in AdS$_4$ \cite{Callan:1989em, Gripaios:2008rg, Rattazzi:2009ux}.  In general, a careful treatment of boundary conditions, such as in \cite{Amsel-Marolf}, is an important additional step which we defer to future work.} Note that all non-vanishing terms in the variation are proportional to $\lam$ since our Lagrangian was constructed to be supersymmetric in the limit $\lam\rightarrow 0$. Requiring that the variation \eqref{eq:LagVar} vanishes for general $\lam$ then yields the conditions\footnote{Another consequence is that the functions $u_{jk,i}$ are symmetric with respect to all their indices. Therefore, the term $(\xi\chi^i)(\chi^j\chi^k) u_{jk,i}$ in \eqref{eq:LagVar} automatically vanishes due to a Fierz identity.}
\be \label{eq:Consusy}
  r = 3(\Sads(\csc) + \bSads(\csc) + \lam K(\csc,\bcsc))\,   , 
  \quad\!\!  u_{ij} = -\tfrac{1}{2} (K_{ij}(\csc,\bcsc) + \Delta_{ij}(\csc))\,  , \quad\!\! t_i = (K_i(\csc,\bcsc) + \Delta_i(\csc))\,  ,
\ee
where $\Delta(\csc)$ is an undetermined holomorphic function and where we have defined the ``\ads\ superpotential'' 
\be \label{eq:defSAds}
  \Sads(\csc) = \Sfl(\csc) + \lam\,\Delta(\csc) \ ,
\ee  
which is convenient since the action can be written only in terms of $\Sads$, with no reference to $P$ or $\Delta$ separately.

Inserting the conditions \eqref{eq:Consusy} into the ansatz \eqref{eq:Lag-ansatz}, we arrive at the supersymmetric \ads\  Lagrangian
\bea
\label{eq:ads4-lag-aux}
\cL(\lam) &=& - g_{i\jb} \p_m \csc^i \p^m \bcsc^\jb - i g_{i\jb}\chib \sigb^m \mathfrak{D}_m \chi^i +g_{i\jb} F^i \Fb^\jb - F^i \Big( \tfrac{1}{2} g_{i\jb,\kb} \chib^\jb \chib^\kb - (\Sads_i + \lam K_i ) \Big)  \nonumber \\
&&  - \Fb^\ib \Big( \tfrac{1}{2} g_{j\ib,k } \chi^j \chi^k - (\bSads_\ib+\lam K_\ib) \Big)  +  \tfrac{1}{4} g_{i\jb,k\lb} \chi^i\chi^k\chib^\jb\chib^\lb - \tfrac{1}{2}(\Sads_{ij} + \lam K_{ij}) \chi^i \chi^j  \bigg. \nonumber \\
&&   - \tfrac{1}{2} ( \bSads_{\ib\jb}+\lam K_{\ib\jb})\chib^\ib \chib^\jb  + 3\lam \bSads + 3\lam \Sads + 3 \lam^2 K \ .
\eea 
The equation of motion for the auxiliary field $F^i$ is
\begin{equation}
g_{i\ib}F^i=( \tfrac{1}{2} g_{\ib m}\Gam^m_{jk} \chi^j \chi^k - (\bSads_\ib+\lam K_\ib) ) \ , \label{eq:F-term}
\end{equation}
so we can integrate it out to obtain
\bea 
\cL(\lam) &=& - g_{i\jb} \p_m \csc^i \p^m \bcsc^\jb - i g_{i\jb}\chib^\jb \sigb^m \mathfrak{D}_m \chi^i    - g^{i\ib}(\Sads_i + \lam K_i )(\bSads_\ib+\lam K_\ib) + 3\lam \bSads + 3\lam\Sads + 3 \lam^2 K  \nonumber \\
&&     - \tfrac{1}{2}\mathfrak{D}_i (\Sads_{j} + \lam K_{j}) \chi^i \chi^j  - \tfrac{1}{2} \overline{\mathfrak{D}}_\ib ( \bSads_{\jb}+\lam K_{\jb}) \chib^\ib \chib^\jb    +  \tfrac{1}{4} \mathcal{R}_{i\jb k\lb} \chi^i\chi^k\chib^\jb\chib^\lb \ . \label{eq:ads4-lag}
\eea 
The scalar potential, which is simply $g^{i\jb} W_i \Wb_\jb$ in the flat space case, is
\be
V(\csc, \bcsc) =  g^{i\ib}(\Sads_i + \lam K_i )(\bSads_\ib+\lam K_\ib) - 3\lam \bSads - 3\lam\Sads - 3 \lam^2 K  \ . \label{eq:ads-scalar-pot}
\ee

The resulting Lagrangians \eqref{eq:ads4-lag-aux} and \eqref{eq:ads4-lag} are not invariant under a K\"ahler transformation of the K\"ahler potential alone, but they are invariant if supplemented with a shift of the \ads\ superpotential:
\be \label{eq:KAds}
  K(\csc,\bcsc) \rightarrow K(\csc,\bcsc) + f(\csc) + \bar f(\bcsc) \ , \qquad
  \Sads(\csc) \rightarrow  \Sads(\csc) - \lam \, f(\csc) \ .
\ee  
As a consequence, the undetermined function $\Delta(\csc)$ in \eqref{eq:Consusy} could be absorbed, through a K\"ahler transformation, into the K\"ahler potential. However, for a particular K\"ahler potential $K$ (chosen in the flat space limit $\lambda\rightarrow 0$), there are distinct choices to extend the supersymmetric flat space Lagrangian to inequivalent supersymmetric \ads\ Lagrangians. For a fixed K\"ahler potential $K$, these choices are distinguished by the holomorphic shift $\Delta$ in the \ads\ superpotential $\Sads$.


\subsection{Supersymmetric Vacua in \ads}

The supersymmetric vacua of this theory, preserving the full AdS$_4$ invariance, satisfy
\be
Q_\alf |0\rangle = 0 \ \Longleftrightarrow \ \langle \delta \chi^i \rangle = 0 \ \Longleftrightarrow \ F^i = 0 \ \Longleftrightarrow \ W_i + \lam K_i = 0 \ . \label{eq:AdS-susy-condition}
\ee
These constitute $n = \dim (X)$ equations for $n$ complex variables and generically result in a \emph{discrete} set of vacua.  Even when $W\equiv 0$, the moduli space of an $\cN=1$ sigma model in \ads\ is {\em not} a copy of the target space $X$, but rather a set of isolated points {\em on} $X$, with the masses of generic scalar fields being proportional to $\lam$.  Since we have assumed that $\lam \ll 4\pi f_\pi$ is an IR scale, the fields are still ``light''.  The conditions for a supersymmetric vacuum in AdS$_4$ were first written down and analyzed in \cite{Ivanov:1979ft} for the Wess-Zumino model, and for general supersymmetric sigma models in \cite{Aharony:2010ay}.

The fact that the would-be moduli are lifted in \ads\ should not come as a surprise.  In gravity, \ads\ is often described as a homogeneous box for gravity, with light traveling to the boundary and back in finite observer time.  Importantly, this is a good description whether the metric is dynamical or not.  The fact that the spectrum of our sigma model is gapped in \ads\ with the moduli space generically reduced to a set of isolated points is simply a reflection of the effectively compact nature of \ads.

Note that the vacuum energy at a supersymmetric vacuum in \ads\ need not vanish. The scalar potential \eqref{eq:ads-scalar-pot} at a supersymmetric vacuum is
\begin{equation}
V|_{susy} = -3\lam (W| + \Wb| + \lam K|) \label{eq:ads-susy-vac-energy} \ ,
\end{equation}
where the quantities on the right hand side are evaluated at the supersymmetric vacuum. Different supersymmetric vacua of the same theory can, in general, have different vacuum energies in \ads. This is another difference from Poincar\'e supersymmetry where all supersymmetric vacua have identical vacuum energies, conventionally taken to be zero. 
This is not a problem since the absolute energy of a state carries no invariant meaning in the absence of gravity. In weakly coupling an \ads\ supersymmetric theory to gravity, however, the vacuum energy of the globally supersymmetric sector results in a tadpole for the metric fluctuation. The AdS$_{4}$ scale receives an $M_{pl}$ suppressed correction,
\begin{equation}
\delta \lam^2 = \frac{V|_{susy}}{M_{pl}^2} \, ,
\end{equation}
which can be disregarded in the limit of weak gravity. Furthermore, in the rigid limit the vacuum energy can be absorbed into a constant shift of the superpotential,
\begin{equation}
W \rightarrow W + \frac{1}{6 \lam} V|_{susy}.
\end{equation}
With this freedom, we can set the vacuum energy of any given supersymmetric vacuum to zero and then weakly couple to supergravity to avoid the gravitational tadpole.

\vspace{0.1in}

\noindent
{\bf Example}: $X=\C \ , \ \ K(\csc,\bcsc) = \bcsc \csc \ .$

\noindent
The Lagrangian in the simple case of a free chiral multiplet with no superpotential is
\be
\cL = -\partial_m \csc\partial^m \bcsc -i\chib \sigb^m\nabla_m \chi + 2 \lam^2 \csc\bcsc \ .
\ee
We see that even in the absence of a superpotential, the complex scalar has a tachyonic mass $m^2 = -2\lam^2$, which is above the Breitenlohner-Freedman bound $-(9/4)\lam^2$ for AdS$_4$ \cite{BF}. In the flat space case, the free complex scalar has a moduli space of vacua parameterized by $\langle\csc\rangle$. In the AdS$_{4}$ case, we see that there is a single vacuum at $\langle\csc\rangle=0$.


\subsection{Target Space Geometry for $\cN=1$ Sigma Models in \ads} \label{sec:TarGeometry}

Thus far we have examined the local structure of supersymmetric \ads\ \smodel{}s, so let us now consider their global properties. In this regard, the K\"ahler transformation \eqref{eq:KAds} of the superpotential has crucial implications on the global structure of the K\"ahler target space geometry $X$: the superpotential $\Sads$ can only be extended to a non-singular holomorphic superpotential on the whole K\"ahler target space $X$ if the target space K\"ahler form $\omega$ is cohomologically trivial, $\left[\omega\right]=0$.\footnote{The easiest way to see this is to note that the combination $K + \lam^{-1}(W + \overline{W})$ is a function on all of $X$ that is also a potential for the K\"ahler form $\omega$. There are therefore two options: first, the K\"ahler form is exact; second, the combination $K+\lam^{-1}(W+\overline{W})$ has branch cuts or singularities, necessarily creating divergences in the scalar potential \eqref{eq:ads-scalar-pot} that effectively decompactify the target space by confining the scalars to the regions away from infinite potential energy.}   
To argue this formally, we cover the K\"ahler target space by holomorphic patches $U_{\alpha}$ decorated with local K\"ahler potentials $K_{\alpha}$ and local (holomorphic) superpotentials $W_{\alpha}$ in each patch. On double overlaps of two patches, $\alpha$ and $\beta$, the holomorphic K\"ahler transition functions obey $\lam f_{\alpha\beta}(\csc)=W_\beta(\csc) - W_\alpha(\csc)$, in  accordance with \eqref{eq:KAds}, implying that the \v Cech cocyle $(\delta f)_{\alpha\beta\gamma}$ vanishes identically on all triple overlaps,
$$
(\delta f)_{\alpha\beta\gamma} = f_{\alpha\beta} - f _{\alpha\gamma} + f_{\beta\gamma}= 0\ .
$$
If the K\"ahler form is non-trivial, the associated \v Cech cocycle $(\delta f)_{\alpha\beta\gamma}$ cannot vanish identically on all triple overlaps \cite{Witten:1982hu}. We thus conclude that the K\"ahler form $\omega$ must be cohomologically trivial for rigid supersymmetric \ads\ \smodel{}s.

\vspace{0.1in}
\noindent
{\bf Example}: $X = \P^1$, Fubini-Study metric.

To illustrate this point, we attempt to construct a supersymmetric \smodel\ in \ads\ with target space $X=\P^1$, endowed with the Fubini-Study K\"ahler metric. Covering the target space $\P^1$ with the two patches $U_0, U_\infty \cong \C$ of the local coordinates $z$ and $w$ that arise from the stereographic projection of $\P^1$, we have the K\"ahler metric and potentials
\be
\begin{aligned}
z\in U_0: \ &&   g_{z\zb} &= \frac{n}{(1+z\zb)^2} =   \partial_z \partial_{\zb}K_0(z,\zb) \ ,
&& K_0(z,\zb) = n \log(1+z\zb) \ , \\
w\in U_\infty: \ &&  g_{w\wb} &= \frac{n}{(1+w\wb)^2} = \partial_w \partial_{\wb}K_\infty(w,\wb) \ , \quad
&& K_\infty(w,\wb) = n \log(1+w\wb) \ . \\
\end{aligned}
\ee
The (non-trivial) K\"ahler form $\omega=i \partial\bar\partial K$ generates $H^{1,1}(\P^1)$ with $\int_{\P^1} \omega = 2\pi n$. On the overlap $U_0\cap U_\infty$ (where $z = w^{-1}$), there is a non-trivial K\"ahler transformation $f_{\infty0}$
\be  
  K_\infty(z^{-1},\zb^{-1}) - K_0(z,\zb) = f_{\infty 0}(z) + \bar f_{\infty 0}(\zb) \ , \qquad f_{\infty 0}(z) = - n \log z = n \log w \ . 
\ee
Due to \eqref{eq:KAds} we know that the superpotential must obey
\be
  W_\infty(z^{-1}) - W_0(z) =  -\lam\, f_{\infty 0}(z) = \lam\,n\log z \ .
\ee
For the local superpotential $W_0(z)\equiv 0$, we arrive at the (bosonic) Lagrangian in terms of the chiral field $z$ in the patch $U_0$
\be
\cL_{U_0 \subset \P^1} = -n \frac{\partial_m z \partial^m \zb}{(1+z\zb)^2} - \lam^2\left(n \ z\zb - 3n \log(1+z\zb)\right) \ ,
\ee
with supersymmetric vacuum for $\langle z\rangle=0$. In the $U_\infty$ patch, with the (singular) superpotential $W_\infty(w) = -n\lam \log w$, the Lagrangian becomes
\be
\cL_{U_\infty \subset \P^1} = -n \frac{\partial_m w \partial^w \zb}{(1+w\wb)^2} - \lam^2\left(\frac{n}{w\wb} + 3n \log\Big(1+\frac{1}{w\wb}\Big)\right) \ .
\ee
Note that due to the infinity of the superpotential $W_\infty$ and the resulting infinity in the scalar potential at $w=0$, a fluctuation in the neighborhood around $w=0$ has arbitrarily high energy. Therefore removing all fluctuations above a certain UV cutoff effectively removes the point $w=0$ (or $z=\infty$) from the target space $X=\P^1$. Thus, the resulting effective \ads\ \smodel\ has the trivialized target space $\C \simeq \P^1\setminus\{w=0\}$ with a cohomologically trivial K\"ahler form $\omega|_{U_0}$.
\vspace{0.1in}


\subsection{Derivation Through Supergravity} \label{sec:SUGRAred}

We have shown that the target space $X$ of a rigid $\cN=1$  \smodel\ in \ads\ has an exact K\"ahler form. Since this is also true for sigma models in supergravity that admit weak-coupling limits \cite{Witten:1982hu, KS}, this gives us an easy route to the \ads\ \smodel\ Lagrangian \eqref{eq:ads4-lag} via a decoupling limit of supergravity.  As we shall now verify, this approach leads to precisely the same theories and the same constraints as found above through a direct analysis of the rigid supersymmetry algebra.

We start with the Lagrangian for supergravity coupled to a set of chiral multiplets that parameterize a K\"ahler target space $X$ (e.g., \cite{Wess:1992cp}). For the present argument it suffices to restrict attention to the bosonic terms of the $\cN=1$ supergravity action, but the derivation goes through for the fermionic terms as well. The general form of the (bosonic) $\cN=1$ supergravity action of interacting chiral multiplets is completely determined in terms of the (local) K\"ahler potential $K(\csc,\bcsc)$ and the superpotential $\Ssugra$
\be \label{eq:SUGRAaction}
\begin{aligned}
  S_{sugra}^{bos}&= -  \int d^4x \sqrt{-\gam}\,     
  \left[\tfrac12 M_{pl}^2 R + g_{i\jb} \partial_m\csc^i\partial^m\bcsc^\jb + V(\csc,\bcsc) \right] \ , \\
  V_{sugra}(\csc,\bcsc)&=e^{K/M_{pl}^2}\left[ g^{i\jb} \left(\Ssugra_i + \frac{K_i}{M_{pl}^2}\Ssugra\right) \left(\bSsugra_\jb + \frac{K_\jb}{M_{pl}^2} \bSsugra\right) - \frac{3}{M_{pl}^2} |\Ssugra|^2 \right] \ ,
\end{aligned}
\ee    
where $M_{pl}$ is the four-dimensional Planck mass. The mass dimensions of the fields are chosen as in \eqref{eq:sig-scaling}, and the K\"ahler potential $K$ and the superpotential $\Ssugra$ have dimensions two and three, respectively. 

The dynamics of the \smodel\ are controlled by the cutoff scale $\Lam_\sig$ and various mass scales $\mu \lesssim \Lam_\sig$ that appear in the superpotential. In order to decouple gravity, we assume that all these scales are much smaller than the Planck scale $M_{pl}$. This prevents, for example, low-energy scattering of \smodel\ fields from producing final state gravitons, as the graviton coupling (via the stress tensor) is suppressed by inverse powers of $M_{pl}$. Graviton self-interactions are also suppressed in the low-energy regime. The dynamics at energies $E \ll \Lam_\sig \lesssim 4\pi f_\pi$ therefore consists of a sector of noninteracting soft gravitons and a decoupled sector of interacting \smodel\ fields. 

The metric satisfies the Einstein equations and is sourced by the stress-energy tensor of the matter sector. If we assume that all the energy scales associated to the \smodel\ sector are small, as we argued in the previous paragraph, the metric equation of motion would yield Minkowski space as a solution. In order to obtain \ads, we introduce a constant term to the supergravity superpotential\footnote{Even in the absence of a matter sector we may introduce a constant supergravity superpotential $P=\lam M_{pl}^2$, describing supergravity in a background with a negative cosmological constant.} 
\be \label{eq:superpot-shift}
\Ssugra_{sugra}(\csc) = \lam M_{pl}^2 +  \Sads(\csc) \ ,
\ee
which gives rise to a negative cosmological constant in the supergravity scalar potential. The quantity $\Sads$ will turn out to be the superpotential that appears in the global supersymmetric \ads\ \smodel\ Lagrangian.
With the addition of the constant term, the scalar potential in an $M_{pl}^{-2}$ expansion takes the form
\be
V_{sugra} = -3 M_{pl}^2 \lam^2 +  \Big( g^{i\jb}(W_i + \lam K_i)(\Wb_\jb + \lam K_\jb) - 3\lam(W+\Wb + \lam K)\Big) + \cO\Big(\frac{1}{M_{pl}^2}\Big)
\ee
The metric equation of motion, to leading order in $M_{pl}$, is
\be \label{eq:EOMmet}
R_{mn} = 3 \lam^2 \, \gam_{mn} \ ,
\ee
which gives \ads\ with the radius $\lam^{-1}$. The terms at order $M_{pl}^0$ above agree precisely with the \ads\ scalar potential in \eqref{eq:ads-scalar-pot}. Moreover, the complete Lagrangian \eqref{eq:ads4-lag}, including all the fermion terms, can be reproduced through the decoupling procedure we have described.

The supergravity action is invariant under K\"ahler-Weyl transformations
\be \label{eq:Ksugra}
K \ \rightarrow \ K + f(\csc) + \bar f(\bcsc) \ , \qquad \Ssugra \ \rightarrow \ \exp(-f(\csc)/M_{pl}^2) \ \Ssugra \ ,
\ee
where the fermions are also rotated by a phase dependent upon $\textrm{Im}(f)$.  Note that the supergravity superpotential $\Ssugra$ transforms as a holomorphic section of a line bundle over the target space \cite{Witten:1982hu}, which is necessary in order for the scalar potential $V(\csc,\bcsc)$ to remain invariant. The modified K\"ahler invariance \eqref{eq:KAds} of the \ads\ Lagrangian can be derived in the gravity decoupling limit from the supergravity K\"ahler invariance above.  Heuristically, we assume that the mass scales of the K\"ahler transformation functions $f(\csc)$, which are of mass dimension two in our conventions, are all associated with the \smodel\ scales and are much smaller than $M_{pl}$. This is a somewhat vague restriction on the class of allowed K\"ahler transformations and it can be made mathematically more precise. Here we avoid presenting all the necessary technical details, as it is intuitive --- due to the mentioned separation of scales --- that the K\"ahler transformation of $\Ssugra = \lam M_{pl}^2 + \Sads$, \eqref{eq:Ksugra}, expanded in powers of $M_{pl}$, yields the AdS$_{4}$ K\"ahler transformations \eqref{eq:KAds}.

\section{Lessons for Constraints in Flat Space and Beyond}

In the previous section, we studied the consistency conditions required for $\cN=1$ supersymmetry of sigma models in AdS$_4$.   It is instructive to consider these results in relation to the consistency conditions for $\cN=1$  sigma models in flat space, as well as to general results on supergravity in four dimensions.  

In flat space, rigid $\cN=1$  supersymmetry does not impose any topological conditions on the cohomology class of the K\"ahler form, nor does the superpotential transform under K\"ahler transformations.  When supersymmetry is gauged, on the other hand, the work of Bagger and Witten \cite{Witten:1982hu} shows that the target space
must have an even integral K\"ahler class $\omega$ in $H^2(X,\mathbb{Z})$ (normalized by $M_{pl}$). Therefore when the K\"ahler form $\omega$ is not exact, the dimensionful irrelevant couplings of the \smodel\ that are associated to the non-trivial holomorphic cycles of the K\"ahler target space are necessarily quantized in units of $M_{pl}$. For example, when $X=\P^1$ the sigma model is characterized by a single dimensionful scale $f_\pi$, so the Bagger-Witten analysis shows that the ratio $\tfrac{f_\pi^2}{M_{pl}^2}$ is an even integer. We cannot, in such a situation, dial the scales $f_\pi$ and $M_{pl}$ independently. 
As a result of this quantization, only Planck-scale experiments would be able to probe the curvature of the $\P^1$ target space, or the irrelevant interactions specific to the $\P^1$ sigma model.  
At energies much lower than the Planck scale, the \smodel\ is essentially trivialized to a local patch $\C\subset\P^1$. Such field theories do not have the interpretation of being weakly coupled to gravity and are intrinsically gravitational.

This raises the interesting question of which rigid supersymmetric field theories can be weakly coupled to gravity.  Komargodski and Seiberg recently 
approached
this question by studying the conditions under which the stress tensor and supercurrent could fit into a single 
flat-space supersymmetry multiplet \cite{KS} (see \cite{Butter:2011ym} for a very recent extension of these arguments to AdS$_{4}$).  
They showed that the standard Ferrara-Zumino multiplet \cite{Ferrara:1974pz}\ is only globally well-defined when the K\"ahler class of $X$, $\left[\omega\right]$, is trivial.  When $\left[\omega\right]\neq0$, one can find a different set of supersymmetric current multiplets that can be coupled to supergravity with an additional linear multiplet.  When the target space $X$ is a Hodge manifold, this additional linear multiplet can be dualized to a chiral superfield such that the enlarged geometry $\hX$ has a trivial K\"ahler class.  The enlarged geometry $\hX$ is then a $\mathbb{C}^*$-fiber bundle over $X$ of the form discussed in Appendix~\ref{sec:fibration}.

As is by now clear, these conditions are all equivalent to the conditions for unbroken $\cN=1$  supersymmetry in an AdS$_4$ background.  Remarkably, these conditions also imply the vanishing of (mixed) gravitational anomalies, even around flat space, as we will shortly explain.  The remainder of this section is devoted to explaining this connection with anomalies and expanding this lesson into a general conjecture about the conditions for a general rigid $\cN=1$  theory to arise as the decoupling limit of some $\cN=1$ supergravity.

\subsection{Anomalies and Constraints on (De-)Coupling Gravity}

As we found in Section \ref{sec:SUGRAred}, the classical conditions for rigid $\cN=1$  supersymmetry in \ads\ spacetimes are equivalent to the conditions for the rigid theory to arise as a decoupling limit of $\cN=1$  supergravity.  However, since the supergravity multiplet contains a gravitino, any such decoupling limit alters the chiral spectrum.  It is thus possible that decoupling (or re-coupling!) gravity, while classically straightforward, is quantum-mechanically obstructed by anomalies in either the global or local supersymmetric theory.   In the remainder of this section, we shall check for such potential obstructions to (de-)coupling gravity and our sigma model by studying the possible mixed-gravitational anomalies in both local and global theories.\footnote{Note that, in general, this analysis should include a careful discussion of potential boundary terms which can have important effects on the chiral spectrum.  For example, it is impossible to find boundary conditions which allow chiral matter to couple to a massless gauge field in AdS$_4$.}

In a globally supersymmetric $\cN=1$ \smodel\ $\csc: \Sigma \rightarrow X$ from a four-dimensional spacetime background $\Sigma$ into a K\"ahler target space $X$, the fermions $\chi$ in the chiral multiplets transform as spinor-valued sections of the pullback tangent bundle $\csc^*TX$. As a consequence, the global $\cN=1$  \smodel\ is anomaly-free when the six-form anomaly polynomial vanishes \cite{Moore:1984ws},
\be
   P_{\rm global}^{(\Sigma,X)} 
   \,=\, \left. \hat A(\Sigma) \wedge \tilde\csc^*{\rm ch}(X) \right|_{\rm (6-form)}
   \,=\, \tilde\csc^*{\rm ch}_3(X)-\frac{1}{24} \tilde\csc^* c_1(X) \wedge p_1(\Sigma) \ .
\ee 
Here ${\rm ch}(X)$ denotes the total Chern character of the target space manifold $X$, and $\hat A(\Sigma)$ and $p_1(\Sigma)$ are the $A$-roof genus and the first Pontryagin class of the spacetime, $\Sigma$, respectively.\footnote{$\hat A(\Sigma)=1-\tfrac{p_1(\Sigma)}{24}$ for four-dimensional manifolds $\Sigma$.} The map $\tilde\csc$ is directly related to the \smodel\ map $\csc$, for a detailed definition of which we refer the reader to \cite{Moore:1984ws}.

For the $\cN=1$ supersymmetric \smodel\ $\csc: \Sigma \rightarrow X$ in four spacetime dimensions coupled to gravity, chiral fermions are spinor-valued sections of the bundle $\csc^*(TX\otimes\mathcal{K})$, where $\mathcal{K}$ is the K\"ahler line bundle obeying $c_1(\mathcal{K}) \simeq \tfrac12 \omega$ and $\omega$ is the K\"ahler form of the target space $X$.\footnote{In $\cN=1$ supergravity, the target $X$ must be a Hodge manifold with an even integral K\"ahler form $\omega$ \cite{Witten:1982hu}.} In addition, the gravitino $\psi_\mu$ transforms as a spinor-valued section of $(T\Sigma \ominus {\bf 1})\otimes\csc^*{\mathcal K}^{-1}$ \cite{Distler:2010zg}.
Therefore the resulting six-form anomaly polynomial for the $\cN=1$ supersymmetric \smodel\ coupled to gravity reads \cite{Distler:2010zg}
\be
\begin{aligned}
  P_{\rm local}^{(\Sigma,X)} 
  \,=\, & \left. \hat A(\Sigma)\wedge \left[ {\rm ch}\,\tilde\csc^*(TX \otimes \mathcal{K})
     -{\rm ch}\left((T\Sigma \ominus {\bf 1})\otimes\csc^*{\mathcal K}^{-1}\right) \right] \right|_{\rm (6-form)}\\[1.5ex]
  \,=\, & 
   \tilde\csc^*{\rm ch}_3(X)-\frac{1}{24}\tilde\csc^* c_1(X) \wedge p_1(\Sigma)+
   \tilde\csc^*c_1(\mathcal K) \left( \tilde\csc^*{\rm ch}_2(X)+\frac{21-n}{24}p_1(\Sigma) \right)\\
   &\qquad\qquad + \frac{1}{2}\tilde\csc^*c_1(\mathcal K)^2\wedge \tilde\csc^*c_1(X)+
   \frac{n+3}6 \tilde\csc^*c_1(\mathcal K)^3 \ .
\end{aligned}
\ee
Here $n$ is the complex dimension of the target space manifold $X$.

The important observation is now that the local anomaly $P_{\rm local}^{(\Sigma,X)}$ decomposes as 
\be
\begin{aligned}
   P_{\rm local}^{(\Sigma,X)} \,=\, & P_{\rm global}^{(\Sigma,X)} + \Delta P^{(\Sigma,X)} \ , \\[2.5ex]
   \Delta P^{(\Sigma,X)} \,=\, & \tilde\csc^*c_1(\mathcal K)
   \left[  
   \left( \tilde\csc^*{\rm ch}_2(X)+\frac{21-n}{24}p_1(\Sigma) \right) \right. \\
   &\left. \qquad\qquad\qquad +  \frac{1}{2}\tilde\csc^*c_1(\mathcal K)\wedge \tilde\csc^*c_1(X)+
   \frac{n+3}6 \tilde\csc^*c_1(\mathcal K)^2  \right] \ ,
\end{aligned}
\ee   
where the contribution $\Delta P^{(\Sigma,X)}$ is proportional to $c_1(\mathcal{K})$ and therefore to the K\"ahler class $\left[\omega\right]$ of the target space $X$. Thus, if the target space $X$ has a cohomologically trivial K\"ahler form, then the process of weakly coupling to gravity does not change the \smodel\ anomaly. In particular, if the global $\cN=1$ \smodel\ is anomaly free then the addition of gravity does not introduce an additional anomaly.

For global $\cN=1$ supersymmetric \ads\ \smodel{}s, we observed in Section \ref{sec:TarGeometry} that the target space K\"ahler form must be cohomologically trivial. Thus if the global $\cN=1$ supersymmetric \ads\ \smodel\ (of chiral mutliplets) is free of anomalies, then there are no further anomaly constraints in coupling to gravity in an \ads\ background.  We conclude then that the vanishing of gravitational anomalies is already guaranteed by the \emph{classical} consistency of these models!


\subsection{The Background Principle}

It is quite a remarkable fact that the analysis of Section \ref{sec:ads4-sigma}, which is completely classical, implies the vanishing of quantum anomalies above, \ie\ that the K\"ahler form must be exact.   There is a simple reason for this: the (mixed) gravitational anomalies tell us the conditions for consistently coupling a microscopic theory to gravity.  Any theory that can be consistently coupled to gravity should also be able to be expanded around a non-trivial metric which solves the equations of motion and preserves the same symmetries.  For this purpose, AdS$_4$ is peculiarly well-suited as it is maximally symmetric, preserves supersymmetry, and arises as a one-parameter deformation of the theory in flat space --- including the supersymmetry algebra and its representation theory.  We thus expect the classical conditions for $\cN=1$  supersymmetry in \ads\ to correspond to necessary conditions for consistently coupling the flat-space theory to supergravity, which is exactly what we found.

Similar effects obtain in other contexts.  For example, consider the bounds on the signs of leading irrelevant operators as discussed in \cite{AdamsNima}.    These bounds can be identified in two ways: when expanding around the trivial vacuum, these constraints can only be seen from a dispersion relation for the quantum $S$-matrix elements; on the other hand, when expanding about a suitable classical background, these constraints are visible classically and at low energies.  Roughly speaking, working in a non-trivial background takes a microscopic (quantum) effect and exponentiates it via multiple scattering off the classical background.

All of this entices us to make a more general conjecture, which we will call the ``Background Principle'': any rigid $\cN=1$ theory in Minkowski space which can be consistently, quantum-mechanically coupled to $\cN=1$ supergravity --- or, conversely, which can arise as the decoupling limit of a well-defined $\cN=1$  supergravity theory --- must also behave smoothly, \emph{as a classical theory}, under a deformation of the rigid Minkowski spacetime to AdS.  In this paper, we have shown this to be the case for conventional sigma models containing only chiral superfields; we conjecture this to be true for all rigid $\cN=1$ theories.

If true, this principle affords both a straightforward route to identifying four-dimensional QFTs which cannot be coupled to supergravity, and leads to a strong statement about the moduli spaces of theories which can be coupled to supergravity and admit a UV completion: their $\lambda\to0$ moduli spaces, governed by long-distance sigma models, must necessarily be non-compact lest supersymmetry be broken when expanding about a rigid AdS$_{4}$ background.

\section{Comments on Moduli Stabilization} \label{sec:ModStab}

We use the $\cN=1$ AdS$_4$ sigma model to study the moduli sector of a large class of string compactifications, concluding, on general grounds, that these compactifications necessarily have moduli whose masses are proportional to the AdS$_{4}$ scale.  Such $\cN=1$ compactifications arise, for example, in the first stages of KKLT/KKLMMT-type scenarios \cite{KKLT,KKLMMT}, where the resulting light moduli can be dealt with by a supersymmetry-breaking uplift.  The lesson of our analysis, which depends only on simple properties of $\cN=1$ sigma models in AdS$_4$, is that such light moduli arise very generally in a model independent fashion, and that lifting them requires either moving away from large volume or breaking supersymmetry, as in the specific scenarios of~\cite{KKLT,KKLMMT}.

We begin with a lightning review of type IIB supersymmetric \ads \, flux vacua (see \cite{Douglas-Kachru-review, Blumenhagen:2006ci, Denef-les-houches} for a general review of various moduli stabilization scenarios).  In the large volume regime of type IIB Calabi-Yau compactifications, the moduli fields can heuristically be divided into the K\"ahler and complex structure moduli of the Calabi-Yau threefold, the complexified axio-dilaton, and the brane moduli. In the presence of spacetime filling D3 branes, the latter moduli include the D3 position moduli in the compactification space. In these large volume scenarios, the complex structure and the axio-dilaton are typically stabilized at weak coupling by turning on R-R and NS-NS fluxes that thread 3-cycles in the Calabi-Yau manifold. These fluxes give relatively large masses to the moduli through the flux-induced superpotential, and they introduce a warp factor \cite{GKP}.\footnote{Strictly speaking, the process of turning on background fluxes does not just introduce a warp factor, but also requires compactification spaces beyond Calabi-Yau manifolds. In the context of type II compactifications to supersymmetric AdS$_4$ vacua, such generalizations are discussed in \cite{Lust:2009zb, Lust:2009mb}.} At this stage, in the sketched approximation, the K\"ahler moduli and the D3 brane moduli are still massless. However nonperturbative effects, such as Euclidean D3 instantons or gaugino condensation on 7-branes, induce a nonperturbative superpotential that can stabilize the remaining moduli and yield an $\cN=1$ \ads \ vacuum. In this scheme, the masses of the D3 brane and K\"ahler moduli (in string units) are exponentially small in the volume of the internal space, while the masses of the complex structure moduli and the axio-dilaton are relatively larger as they depend on the volume through an inverse power law.

Scenarios of moduli stabilization that make use of the supergravity approximation are consistent in the limit that the volume of the internal space and all of its cycles are large in string units. This is precisely the gravity decoupling limit discussed in earlier sections since the four-dimensional Planck mass is related to the volume of the internal space in string units as
\be
M_{pl}^2 = \frac{\rm{Vol}}{g_s^2} M_s^2\, ,
\ee
Vol $\rightarrow \infty$ thus implies that $M_{pl}/M_s \rightarrow \infty$.\footnote{Note that we cannot strictly take this limit since then the internal space decompactifies.} In this limit, the light moduli can be modeled as a supersymmetric sigma model in \ads. Another consequence of the large volume limit is that only the leading nonperturbative effects appear in the superpotential. 
We will argue that for a generic supersymmetric AdS flux vacuum at large volume (in type IIB scenarios), the masses of the light moduli are all proportional to $\lam$. Our conclusions clearly do not apply to the 
scenarios discussed in \cite{LVS}, since the AdS minima in those constructions are non-supersymmetric. For a detailed analysis of the moduli spectrum in these models, see \cite{LVS-moduli}.

The structure of this section is as follows: first, we consider an example with a single K\"ahler modulus, then we consider an example with additional brane moduli. We point out that the mass matrix of the light modes in these scenarios is proportional to $\lam^2$, and we end with a general argument that shows the existence of light moduli even when one allows for multiple K\"ahler moduli.

\vspace{0.1in}
\noindent
{\bf Example 1} (K\"ahler modulus): $K = - \log(y + \yb) \ .$

This is the sigma model encountered in the case of a single K\"ahler modulus, where $\rho \equiv y+\yb$ measures the volume of the internal space. (We set $M_{pl}=1$ throughout this section.) The imaginary part $a\equiv \Im(y)$ is an axion that enjoys a continuous shift symmetry. The supersymmetric vacua of the theory are determined by solving \eqref{eq:AdS-susy-condition}, which yields a supersymmetric vacuum at $\rho \rightarrow \infty$. This runaway is typical of the no-scale structure where the scalar potential vanishes as the internal space decompactifies. Of course, $y$ can be naturally stabilized at large volume by including a ``small'' superpotential term. There is, however, a more general argument for a superpotential term --- since this sigma model is part of a consistent quantum gravity theory, the continuous shift symmetry $a \rightarrow a + c$ must be broken by non-perturbative effects. This motivates a superpotential term $W(y) = \mu^3 \exp[-y]$. The negative exponential ensures that the superpotential vanishes, as it should, in the limit of large volume. Solving the supersymmetry conditions \eqref{eq:AdS-susy-condition}, we find a single supersymmetric vacuum at $\rho=\rho_0$ and $a=\pi+3\arg(\mu)$, where
\be
\rho_0 \exp (-\rho_0/2) = \frac{\lam}{|\mu|^3} \label{eq:rho-vev} \ .
\ee
Since the superpotential scale $\mu \lesssim \Lam_\sig \ll M_{pl}=1$, this suggests that a supersymmetric vacuum exists at large $\rho_0$ if and only if the AdS scale $\lam$ is exponentially small (in units of $M_{pl}$). The masses of fluctuations of $\rho$ and $a$ can be easily computed and are both $\approx \rho_0^2 \lam^2$. Note that the masses are proportional to $\lam$ and hence these moduli are light.\footnote{Since $\rho_0 \gg 1$, however, the masses of the light moduli are parametrically larger than the AdS scale.} The superpotential could, in general, include higher order exponentials $\exp(- n y)$ for $n \in \Z^+$. By working in the large volume limit, we can drop these terms and keep only the leading exponential.

\noindent
As an example, for $\rho_0 = 50$ and $\mu = 1$, we find 
\be
\frac{\lam}{M_{pl}} \approx 7 \times 10^{-10} \ , \qquad m_a^2 = 2350 \lam^2 \ , \qquad m_\rho^2 = 2448 \lam^2 \ .
\ee

\vspace{0.1in}
\noindent
{\bf Example 2:} Mobile D3 brane + K\"ahler modulus\,.

This sigma model describes the coupling of the D3 brane position to the overall volume modulus in a IIB compactification. The K\"ahler potential for such a scenario was first written down in \cite{DeWolfe-Giddings} and derived in \cite{Camara:2003ku,Grana:2003ek}:
\be
\hK (y, \yb, z^i, \zb^\ib) = - \log \big[y+\yb - f^2 k(z^i,\zb^\ib)\big] \ , \label{eq:total-Kahler-log}
\ee
where $k(z^i, \zb^\ib)$ is the K\"ahler potential on the brane moduli space, $X$.  In the above formula, we set $M_{pl}=1$ and take $f \ll 1$, which should be thought of as setting the KK scale \cite{D3-pot-1}.  We denote the space spanned by the coordinates $y$ and $z^i$ by $\hX$. This space is a $\C^*$ fibration over $X$ \cite{KKLMMT} with an exact K\"ahler form derived from \eqref{eq:total-Kahler-log}. 

We discuss in detail various aspects of the geometry of $\hX$ in Appendix \ref{sec:fibration}. There we show that such a $\C^*$ fibration over a compact space $X$ is allowed only when it is a Hodge manifold, \ie\ when the (normalized) K\"ahler class is an integral class in $H^2(X,\Z) \cap H^{1,1}(X,\C)$. When this is not the case, there are at least two ways in which one can create a total space $\hX$ with an exact K\"ahler form: first, we could couple the sigma model on $X$ to a single linear multiplet and avoid the quantization condition mentioned above; second, we could add multiple $\C^*$ fibers, up to $h^{1,1}(X)$ of them, to trivialize the K\"ahler form of the total space of the fibration.

When $X$ has a non-trivial K\"ahler form, Komargodski and Seiberg used an analysis of supercurrent multiplets in Minkowski space to show that it is impossible to stabilize the K\"ahler modulus while leaving the brane moduli massless \cite{KS}.  This followed from the fact that the theory with target space $\hX$ has an exact K\"ahler form and hence a well-defined Ferrara-Zumino multiplet.  Then if $y$ became massive while the $z^i$ remained massless, we could integrate out $y$ and be left with a sigma model on $X$.  Since the K\"ahler form on $X$ is non-trivial, the resulting supercurrent multiplet would then not be well-defined over the moduli space, but it is not possible for an RG flow to take well-defined operators from the UV to ill-defined operators in the IR.  Thus, the inclusion of a nonperturbative superpotential would necessarily have to lift both the K\"ahler modulus and the brane moduli.  This result matches with the original stringy arguments of \cite{KKLMMT}.  We will be led to similar conclusions in AdS$_4$ sigma models by simply appealing to the conditions necessary for the existence of supersymmetric vacua at large volume.

In analogy with the previous example, the analysis will be carried out in terms of the variables
\be
\rho \equiv y+\yb-f^2 k(z^i, \zb^\ib) \ , \qquad a \equiv \Im(y) = \frac{y-\yb}{2i} \ .
\ee
As explained in \cite{Camara:2003ku,Grana:2003ek}, the variable $\rho$ measures the overall volume of the space $X$.  In the large volume limit $\rho \rightarrow \infty$, it is consistent to include only the leading exponential in the nonperturbative effective superpotential,
\begin{equation}
W(y, z^i)  =  p(z^i) \exp(-n y) \ .
\end{equation}
As in the previous example, the superpotential breaks the shift symmetry of $a$, but since $y$ is not a good coordinate on the total space $\hX$, there is a $z^i$ dependent prefactor \cite{Witten-nonpert}. $p(z^i)$ is a holomorphic section of a line bundle on $X$ with transition functions chosen precisely to cancel those of the $\C^*$ section $\exp (-n y)$. The resulting $W(y, z^i)$ is simply a function on $\hX$.

The conditions for a supersymmetric vacuum in AdS$_4$ \eqref{eq:AdS-susy-condition}  are
\begin{align}
n W  = - \frac{\lam}{\rho}, \quad 
\frac{p_i}{p} W  =  \frac{\lam f^2 k_i}{\rho} \, , \label{eq:moduli-vevs}
\end{align}
where the subscript $i$ denotes a derivative with respect to $z^i$. Substituting for $W$ in the second equation using the first, we have 
\be
n p_i(z) + f^2 k_i(z,\zb) p(z) =0 \ . 
\ee
The vacuum expectation values for the $z^i$ are completely determined by this equation, independent of $\rho$ and $\lam$. The first equation in \eqref{eq:moduli-vevs} then has a solution with large $\rho$ only when $\lam$ is exponentially small.

The moduli masses can now be computed from the two-derivative matrix of the scalar potential evaluated at the supersymmetric vacuum,
\be\begin{aligned}
V_{ab} & = -\lam (W_{ab} + \lam K_{ab}) \ , \\
V_{a\bb} & = g^{c\db} (W_{ac}+\lam K_{ac})(\Wb_{\bb \db} + \lam K_{\bb \db}) - 2 \lam^2 g_{a\bb} \ , \\
V_{\ab\bb} & = -\lam(\Wb_{\ab \bb} + \lam K_{\ab\bb}) \ .
\end{aligned}\ee
The indices $a,b,\cdots$ run through the coordinates $(y, z^i)$. This mass matrix is proportional to $\lam^2$ if $W_{ab}$ is proportional to $\lam$ for all $a,b$. Here we find
\be
W_{yy}  = n^2 W = -\frac{n \lam}{\rho} \ , \qquad
W_{yi}  = -n \frac{p_i}{p} W = \frac{n\lam f^2 k_i}{\rho}  \ , \qquad
W_{ij}  = \frac{p_{ij}}{p} W = -\frac{\lam}{n\rho }\frac{p_{ij}}{p} \ ,
\ee
which indeed shows that $W_{ab}$ is proportional to $\lam$.  The moduli masses are obtained by canonically normalizing the kinetic terms for the fluctuations and then computing the eigenvalues. The eigenvalues are clearly proportional to $\lam$, but the question is whether the prefactor, which depends on the vevs of $y,z^i$, can modify this scaling. We noted earlier that the vevs of the $z^i$ are completely independent of $\lam$. The vev of $\rho$ does depend on $\lam$, but only in a logarithmic manner, and so we expect the moduli masses to be exponentially small (in the volume). 

\noindent
{\bf General Argument}

This argument for the appearance of light fields easily generalizes. To this end, let us assume that at the classical/perturbative level the analyzed scenarios have a set of shift symmetries $\Im(y^A) \rightarrow \Im(y^A) + const$ with respect to the moduli fields $y^A$, and we denote the remaining moduli in the theory by $z^i$. In type II compactifications, for instance, the  moduli fields $y^A$ could arise from complexified K\"ahler moduli while the fields $z^i$ may represent complex structure moduli of the internal Calabi-Yau spaces. Nonrenormalization theorems for the superpotential severely constrain the form of the effective superpotential such that no perturbative contributions to the effective superpotential can break these shift symmetries. However, there still may be nonperturbative corrections that appear as exponentials $\exp(-n y^A)$ in the effective superpotential. As a consequence, the leading order terms of the (nonperturbatively generated) effective superpotential take the form
\be \label{eq:AdSGeneral}
W(y,z) = \sum_A C^A(y^A,z) = \sum_A p^A(z) \exp( - y^A) \ .
\ee
Here we have absorbed any numerical factors in the exponential into $y^A$. To avoid runaways, we require that the $p^A$, which are general functions of the $z^i$, do not vanish identically for any $A$. The K\"ahler potential $K(y, \yb,z,\zb)$ is such that $y^A$ appear in the combination $y^A + \yb^{\bar{A}}$ so that the shift symmetry is only broken by the superpotential $W$. In addition, we require that the boundaries $y^A \rightarrow \pm \infty$ of the target space are at infinite distance with respect to the K\"ahler metric, a condition that we discuss further in Appendix~\ref{sec:fibration}.

Before we move on, let us point out a few caveats to the arguments that led us to \eqref{eq:AdSGeneral}: first, for general AdS theories we cannot unambiguously identify a unique superpotential since a K\"ahler transformation can shift (part of) the superpotential into the K\"ahler potential and vice versa; however, when our theory has a well-defined flat space limit, $\lambda\to0$, we \emph{can} unambiguously define the superpotential as the surviving part in the limit $\lambda\to0$.  Second, in arguing for the structure of the superpotential we relied upon nonrenormalization theorems for the superpotential, but it is not clear to what extent such theorems are applicable in the context of the global AdS \smodel{}s; again, we are simply guided by the limit $\lambda\to0$ and the intuition gained from phenomena in supersymmetric gauge theories \cite{Intriligator-Seiberg}.  In further defense of the form \eqref{eq:AdSGeneral}, note that it also agrees with the expected structure of nonperturbative effects that arise in string compactifications \cite{Witten-nonpert}. How far such a naive analysis can be pushed in the general AdS$_4$ setting is an interesting question to which we hope to return elsewhere.

The conditions to have a supersymmetric vacuum \eqref{eq:AdS-susy-condition} read
\be
C^A = \lam K_A\ , \qquad \sum_A C^A_i + \lam K_i = 0 \ .
\ee
The subscripts denote derivatives with respect to the corresponding fields. The second derivatives of the superpotential $W$ evaluated in a supersymmetric vacuum are given by
\be
\begin{aligned}
  &W_{AA} \big|_{\textrm{susy}}  =  C^A\big|_{\textrm{susy}} = \lam K_A\big|_{\textrm{susy}}  \ ,   &&W_{AB}\big|_{\textrm{susy}}=0 \quad \textrm{ for } A \neq B \ , \\
  &W_{Ai}\big|_{\textrm{susy}}   =  -C^A_i\big|_{\textrm{susy}}= -\lam \frac{p^A_i}{p^A} K_A\big|_{\textrm{susy}} \ , \quad
  &&W_{ij}\big|_{\textrm{susy}}  =  \sum_A C^A_{ij}\big|_{\textrm{susy}} = \lam\sum_A  \frac{p^A_{ij}}{p^A} K_A \big|_{\textrm{susy}}\ .
\end{aligned}
\ee
The argument is virtually identical as in the previous example. As a result, the moduli masses are proportional to $\lam$ with a prefactor that depends only logarithmically on $\lam$.

\noindent
In summary, we find that for KKLT-like scenarios of large-volume moduli stabilization with $\cN=1$ \ads\ supersymmetry, there are always light moduli with masses proportional to the AdS scale.


\section{Gauge Theories: Regulating the Affleck-Dine-Seiberg Runaway in AdS}

So far, we have focused on $\cN=1$ sigma models in AdS$_{4}$, their consistency conditions, and their implications for certain moduli stabilization scenarios.  
It would be interesting to extend this analysis to $\cN=1$ gauge theories and their moduli spaces, as well.   
While a systematic treatment is beyond the scope of this paper,\footnote{A complete analysis must take into account boundary conditions, which have a significant effect in AdS$_{4}$ \cite{Aharony:2010ay}, and a more thorough treatment of possible corrections to the K\"ahler potential.} our analysis above suggests some interesting predictions on the moduli spaces of $\cN=1$ gauge theories in AdS$_{4}$. For example, placing $N_{f}<N_{c}$ SQCD in AdS$_4$ can regulate the Affleck-Dine-Seiberg runaway \cite{Affleck-Dine-Seiberg}.  In this section we present a schematic analysis of this system. We hope to return to a more detailed discussion of gauge theories in AdS$_4$ in the future.

It was suggested by Callan and Wilczek \cite{Callan:1989em} that AdS$_{4}$ serves as an infrared regulator for theories that would otherwise have incurable divergences in flat space. They analyzed the case of the XY model in AdS$_2$ where the properties of the high temperature phase, in which the vortices are deconfined, can be calculated reliably in a dilute gas approximation. They also proposed that the confining phase of QCD could be studied at weak coupling in AdS$_4$ since the usual IR divergences associated with nonperturbative computations would be regulated.

In this paper, we have studied the Lagrangian for interacting chiral multiplets in AdS$_4$. In some cases, an asymptotically free gauge theory at energies below the confinement scale, $\Lam_c$, can be described by an effective Lagrangian consisting of the bound state mesons and baryons (as in the case of QCD). For example, consider the case of four-dimensional $\cN=1$ $SU(N_c)$ SQCD with $N_f$ quarks in the fundamental representation. The dynamics of the gauge theory depends on the ratio $N_f/N_c$ (see \cite{Intriligator-Seiberg}); in the case $N_f < N_c$, an effective superpotential is generated \cite{Affleck-Dine-Seiberg}
\be
W_{\mathit{eff}} = (N_c-N_f) \left( \frac{\Lam_c^{3N_c-N_f}}{\det M} \right)^{\frac{1}{N_c-N_f}} \ .
\ee
Here $M$ is an $N_f\times N_f$ matrix of meson superfields. As explained in \cite{Intriligator-Seiberg}, this superpotential does not violate nonrenormalization theorems since it is generated by non-perturbative effects (instantons when $N_f=N_c-1$ and gaugino condensation in the unbroken $SU(N_c-N_f)$ gauge group when $N_f < N_c-1$). The classical theory has a moduli space of vacua along which various mesons acquire expectation values, thus Higgsing the gauge group. Quantum mechanically, the entire moduli space is lifted and the theory has no vacuum; instead, it has a runaway $M \rightarrow \infty$.

We can study the behavior of this theory in globally supersymmetric AdS$_4$ by arranging a hierarchy of scales $\Lam_c \gg \lam$, where $\lam$ is the inverse AdS$_{4}$ radius. The space is effectively Minkowski at the confinement scale, so we expect that the effects of the AdS$_{4}$ curvature are negligible at those energies (we will discuss corrections below). For convenience, we specialize to the case $N_f=1,~N_c=2$. Then in terms of the dimensionless meson field $m \equiv M/\Lam_c^2$, the condition for a supersymmetric vacuum reads
\be
\frac{\partial W}{\partial m} + \lam \frac{\partial K(m, \mb)}{\partial m} = 0 \ .
\ee
In contrast to the supersymmetry conditions in Minkowski space, the existence of a supersymmetric vacuum depends crucially on the K\"ahler potential in \ads. 
For example, if we assume that the K\"ahler potential remains canonical (more on this below), $K = 2 \Lam_c^2  |m|$, we find a single supersymmetric vacuum located at
\be
\langle m \rangle = \left( \frac{\Lam_c}{\lam}\right)^{1/2} \gg 1 \ .
\ee
The masses of the meson fluctuations around the supersymmetric vacuum, which can be computed from the scalar potential, are proportional to $\lam$.  Since $\lam \ll \Lam_c$, we can treat the mesonic fields as light propagating degrees of freedom below the confinement scale, so the Affleck-Dine-Seiberg runaway can apparently be regulated by the AdS$_{4}$ scale.

Of course, this discussion ignores boundary conditions and quantum corrections to both the K\"ahler potential and the superpotential.  The effects of boundary conditions are quite subtle and beg for further study, which we leave for future work; for now we will simply assume their effects are negligible in the limit $\lam/\Lam_c \rightarrow 0$.  We have argued that in AdS there is no invariant distinction between the superpotential and the K\"ahler potential. As a consequence, it is difficult to say whether the vacuum we have found is stable against quantum corrections without a detailed calculation. A correction to the K\"ahler potential, for example, of the form  $\Lam_c^j \lam^k |M|^{1-j/2-k/2} = \Lam_c^2 |m| \times \left[\big(\tfrac{\lam}{\Lam_c}\big)^k |m|^{-j/2-k/2}\right]$ for $j,k\in\mathbb{Z}$ and $j,k\geq 0$ does not destabilize the vacuum. Logarithmic corrections, on the other hand, are potentially dangerous if the coefficients are sufficiently large. In this note, we do not further examine the structure of such quantum corrections, but we hope that the presented arguments serve as a motivation to study these theories in greater detail.

Our understanding of supersymmetric gauge theories in flat space was greatly advanced by the holomorphy arguments pioneered by Seiberg \cite{Seiberg-holomorphy}. In the AdS$_4$  case, as we have seen, these arguments can become quite subtle because of the mixing between the superpotential and the K\"ahler potential under K\"ahler transformations. An extremely important task, then, is to understand how AdS$_4$ modifies the rich phase structure of nonabelian gauge theories in Minkowski space.  We hope to return to this important problem in future work.

\vspace{0.15in}
\noindent
{\bf \large Acknowledgements:}

We thank 
J. Distler,
T. Faulkner,
S. Franco,
D. Freedman,
S. Giddings,
Z. Komargodski,
J. Louis,
D. Marolf,
J. Polchinski,
A. Van Proeyen,
D. Robbins,
E. Silverstein,
C. Tamarit,
J. Thaler, 
D. Tong,
G. Torroba,
and
S. Trivedi,
for useful discussions and correspondences.  We would especially like to thank S. Kachru, L. McAllister and N. Seiberg for enlightening discussions,  and S. Kachru for helpful comments on the draft.
A.A. thanks the 2010 Amsterdam Summer Workshop for an enlightening setting in which this work was initiated, and the Stanford Institute for Theoretical Physics and the Tata Institue of Fundamental Research for hospitality while this work was being completed. 
V.K. thanks the organizers of the Advanced String School 2010 in Puri, India, where a part of this research was carried out. 
A.A. is supported by the DOE under contract \#DE-FC02-94ER40818. H.J., V.K., and J.M.L., are supported by the Kavli Institute for Theoretical Physics and in part by the NSF Grant PHY05-51164. H.J. is also supported in part by the Stanford Institute for Theoretical Physics and the NSF Grant 0244728. J.M.L is also supported in part by the NSF Grant No. PHY07-57035. 

\bigskip

\appendix
\section{Geometry of the Enlarged Target Space $\hX$} \label{sec:fibration}

Since the global target space structure of supersymmetric \ads\ \smodel{}s require us (even without coupling to gravity) to consider target space geometries with trivial K\"ahler forms, we discuss next in some detail how to obtain an enlarged target space $\hX$ that lacks compact holomorphic cycles (suitable for an \ads\ \smodel) from a target space geometry $X$ that contains compact holomorphic cycles. In this process, we are naturally led to the same target space enlargement $\hX$ as discussed in \cite{KS}. Hence our results for the K\"ahler target space apply equally well for both scenarios.

Consider an arbitrary K\"ahler manifold $X$ with a non-trivial K\"ahler form $\omega$.\footnote{In this work we consider smooth target space manifolds. Various aspects of singular target spaces are discussed in \cite{Hellerman:2010fv}.} In a patch $U_\alf \subset X$, we can define the K\"ahler potential $k_\alf( \csc, \bcsc)$, where the $\csc$ are holomorphic coordinates in the patch. On the intersection of two patches, the K\"ahler potential undergoes a K\"ahler transformation specified by the holomorphic function $f_{\alpha\beta}(\csc)$
\be \label{eq:Kdouble}
  k_\alpha(\csc,\bcsc) - k_\beta(\csc,\bcsc) = f_{\alpha\beta}(\csc) + \bar f_{\alpha\beta}(\bcsc) \ .
\ee
This determines the $f_{\alpha\beta}(\csc)$ up to imaginary constants
\be \label{eq:famb}
  f_{\alpha\beta}(\csc) \sim f_{\alpha\beta}(\csc) + 2\pi i\, c_{\alpha\beta} \ .
\ee
Moreover, due to the relation \eqref{eq:Kdouble} the transformation functions $f_{\alpha\beta}(\csc)$ must obey
$$
  f_{\alpha\beta}(\csc)-f_{\alpha\gamma}(\csc)+f_{\beta\gamma}(\csc)+
  \bar f_{\alpha\beta}(\bcsc)-\bar f_{\alpha\gamma}(\bcsc)+\bar f_{\beta\gamma}(\bcsc)= 0 \ .
$$  
Then on triple overlaps, we can define the real constants
\be \label{eq:omegaCheck}
  \omega_{\alpha\beta\gamma} \,=\, \frac{1}{2\pi i} \left( f_{\alpha\beta}(\csc)-f_{\alpha\gamma}(\csc)+f_{\beta\gamma}(\csc) \right) \ .
\ee 
These real constants $\omega_{\alpha\beta\gamma}$ are defined modulo the real constants $(\delta c)_{\alpha\beta\gamma}\,=\, c_{\alpha\beta} - c_{\alpha\gamma} + c_{\beta\gamma}$, according to the ambiguity~\eqref{eq:famb}, and they obey $\omega_{\alpha\beta\gamma} - \omega_{\alpha\beta\delta} + \omega_{\alpha\gamma\delta} - \omega_{\beta\gamma\delta}=0$. Thus the constants~$\omega_{\alpha\beta\gamma}$ furnish an element in {\v C}ech cohomology group $\check H^2(X,\mathbb{R})$, which may be identified (by the {\v C}ech-de Rham-isomorphism) with the non-exact K\"ahler $(1,1)$-form $\omega$. 

We construct a new space $\hX$ as a fibration over $X$ in the following way: for each $U_\alf$ we add a fiber coordinate $y_\alf \in \C$. On the overlap of two patches, the local coordinates $y_\alf$  are related through the transition functions $f_{\alf\beta}(\csc)$ as
\be \label{eq:fiber-xform}
y_\alf - y_\beta = f_{\alf\bet}(\csc) \ .
\ee
Since $X$ has a non-trivial K\"ahler form, we have to make sure that these transformations are consistent on triple overlaps: 
\be
y_\alf - y_\bet = (y_\alf - y_\gam) - (y_\bet - y_\gam) = f_{\alf\gam}(\csc) - f_{\bet\gam}(\csc) = f_{\alf\bet}(\csc) - 2\pi i \omega_{\alf\bet\gam} \, .
\ee
Then we see that $y_\alf$ is a good fiber coordinate on the triple overlap only if we identify $y_\alf \sim y_\alf + 2\pi i \omega_{\alf\bet\gam}$. For the chiral field $y$ to be a well-defined periodic field, and hence to ensure a geometric interpretation in terms of a fibration, we require that the constants $\omega_{\alpha\beta\gamma}$ are all mutually commensurate --- \ie, there exists a constant $\kappa \in \R$ for which all the $\kappa\,\omega_{\alpha\beta\gamma}$ are integers.\footnote{When this condition is not satisfied, there is no geometric interpretation of this space since the fiber is periodically identified on a dense set. From the physics point of view, at least in Minkowski space, there is a sensible dual interpretation in terms of a linear multiplet coupled to a \smodel\ with target space $X$. Only when a Dirac quantization condition --- \ie, when the identifications are integrally related --- is obeyed are we allowed to dualize linear and chiral multiplets \cite{Adamietz:1992dk}.} Then we get a periodic chiral field $y$ with periodicity
\be \label{eq:phiper}
  y \sim y + 2 \pi i\,\kappa \ .
\ee
The requirement that we obtain a well-defined periodic chiral field $y$ transforming as in \eqref{eq:fiber-xform} imposes a geometric condition on the K\"ahler target space geometry $X$. Namely, the K\"ahler form $\omega$ needs to be quantized with respect to some positive real constant $\kappa$,
\be \label{eq:Hodgecon}
  \kappa\,\omega \,\in\, H^{1,1}(X)\cap H^2(X,\mathbb{Z}) \ .
\ee
Such K\"ahler manifolds with integral K\"ahler forms are called K\"ahler manifolds of restricted type or Hodge manifolds (see, e.g., \cite{Griffiths:1994}).  (Again, the Hodge condition follows from an attempt to trivialize the K\"ahler class of $X$ by fibering a single line over it.  The more generic situation is discussed in a paragraph below.)

From \eqref{eq:Kdouble} and \eqref{eq:fiber-xform}, we can define a \emph{global} real coordinate $\rho_\alf \equiv y_\alf + \yb_\alf -k_\alf(\csc, \bcsc)$ since the transformations on the overlaps precisely cancel (we henceforth drop the $\alf$ label on $\rho$). Using this coordinate, we can construct a globally well-defined K\"ahler potential on the enlarged space $\hX$ consisting of the coordinates $\csc^i, y,$
\be \label{eq:Khat}
  \hat K(\csc,\bcsc,y,\bar y) = - H\left( \rho \right) = -H( y+\bar y - k(\csc,\bar \csc) ) \ .
\ee
Note that this ansatz for the K\"ahler potential respects the continuous shift symmetry $y \rightarrow y + 2\pi i \,c$. This indicates that there is a dual formulation in terms of a linear multiplet by dualizing the chiral multiplet $y$ \cite{Adamietz:1992dk}. We obtain a positive definite K\"ahler metric $\hat G$, represented by the line element
\be \label{eq:Khat-metric}
  ds_{\hat X}^2 = - H''(\rho) (dy- k_i(\csc,\bar\csc) d\csc^i)(d\bar y - k_{\bar\jmath}(\csc,\bcsc)d\bcsc^{\bar\jmath})
  + H'(\rho) g_{i\bar\jmath}(\csc,\bcsc)d\csc^id\bcsc^{\bar\jmath} \ ,
\ee
as long as $H$ (in a domain of $\rho$) is a smooth real function obeying
\be \label{eq:ConH}
  H'(\rho)>0 \ , \qquad H''(\rho)<0 \ .
\ee

There are a few remarks in order before we further analyze the geometry. For a compact K\"ahler target space $X$ with an integral basis $\omega_A$ of non-trivial $(1,1)$-forms, the Hodge condition \eqref{eq:Hodgecon} imposes that the (rescaled) K\"ahler form $\kappa\,\omega$ is an integral linear combination of the integral basis $\omega_A$. When $H^{1,1}(X)$ is generated by a single non-trivial $(1,1)$-form, we can always rescale the K\"ahler form $\omega$ with an appropriate constant $\kappa$ to achieve integrality. When $\dim(H^{1,1}(X)) > 1$, the Hodge condition is not met for K\"ahler forms arising as linear combinations with irrational coefficients relative to each other. In such a situation, we can trivialize the K\"ahler form by introducing additional periodic chiral fields. For instance, we could, following \cite{Banks-Seiberg}, introduce a periodic chiral field for each $(1,1)$-form $\omega_A$ to compensate its contribution to the K\"ahler form $\omega$. Then the failure of the Hodge condition would be reflected in the fact that at least two of the introduced periodic chiral fields would have periodicities that are irrational relative to each other.  Of course, this is exactly what one must do at a generic point in the K\"ahler moduli space of $X$.

In order to now exhibit the geometric structure of the constructed K\"ahler target space $\hat X$ with the K\"ahler potential \eqref{eq:Khat}, we exponentiate the chiral field $\phi$ and introduce the reparametrized chiral field $U$
\be
   U = \exp\left(\frac{y}{\kappa} \right) \ .
\ee
Due to \eqref{eq:phiper}, the chiral field $U$ yields a local single valued $\mathbb{C}^*$ coordinate in each local patch. Therefore, the constructed target space $\hat X$ is a $\mathbb{C}^*$-fiber bundle over the original target space $X$,
\be \label{eq:Xhatfib}
   \mathbb{C}^* \longrightarrow \hat X \xrightarrow{\ \pi\ } X \ .
 \ee
Note that under K\"ahler transformations $\phi\rightarrow\phi+f(A)$ of the base, the $\mathbb{C}^*$-fiber coordinate $U$ transforms as
 \be
   U \rightarrow g(\csc) U \ , \qquad g(\csc)= \exp\left(\frac{f(\csc)}{\kappa} \right) \ .
 \ee
Here the holomorphic transition functions $g(\csc)$ are locally non-vanishing. We can think of the $\mathbb{C}^*$-bundle as arising from a complex line bundle with its zero section removed. Therefore, analogously to line bundles,  we characterize the $\mathbb{C}^*$-bundle by its first Chern class 
\be
  c_1(\mathbb{C}^*) =  \left[ \kappa\,\omega \right] \in H^{1,1}(X)\cap H^2(X,\mathbb{Z}) \ ,
\ee
which by construction is equal to the integral cohomology representative of the rescaled K\"ahler $(1,1)$-form $\kappa\,\omega$.\footnote{The transition functions $g_{\alpha\beta}(\csc)$ give rise to the first Chern class via the map $H^1(X,\mathcal{O}^*)\rightarrow H^2(X,\mathbb{Z})$ arising from the long exact sequence of the exponential exact sequence of sheaves $0 \rightarrow 2\pi i \,\mathbb{Z} \hookrightarrow \mathcal{O}\xrightarrow{\exp} \mathcal{O}^*\rightarrow 0$ \cite{Griffiths:1994}. Under this map, one finds the $\kappa$-rescaled {\v C}ech representative $\kappa\,\omega_{\alpha\beta\gamma}$ of \eqref{eq:omegaCheck} for the first Chern class.} 

Let us pause to analyze how the $\mathbb{C}^*$-fiber bundle structure \eqref{eq:Xhatfib} of the K\"ahler target space $\hat X$ manages to trivialize its K\"ahler form $\hat\omega$. First of all, a straightforward calculation shows that a globally well-defined trivializing one form $\hat\theta$ exists, with $\hat\omega=d\hat\theta$:
\be \label{eq:theta}
\begin{aligned}
  \hat\theta &= \frac{i}{2}H'(\rho)\left(\kappa \tfrac{dU}{U} - \kappa \tfrac{d\bar U}{\bar U} - 
  \left(K_i(\csc,\bcsc)d\csc^i - K_{\bar\jmath}(\csc,\bcsc)d\bcsc^{\bar\jmath} \right) \right)\\
   &= \frac{i}{2}H'(\rho)\left(d\phi - d\bar\phi  - 
  \left(K_i(\csc,\bcsc)d\csc^i - K_{\bar\jmath}(\csc,\bcsc)d\bcsc^{\bar\jmath} \right) \right)  \ .
\end{aligned}  
\ee  
The relationship between the cohomology classes of the base space $X$ and the whole space $\hat X$ sheds more light on the exactness of the K\"ahler form $\hat\omega$. We can compute the cohomology classes of the $\mathbb{C}^*$-fibration $\hat X$ with the Leray spectral sequence \cite{Bott:1982}. In particular, we find that the two-form class associated to the first Chern class of the $\mathbb{C}^*$-bundle is removed from the two-form cohomology of $\hat X$ since it becomes exact,\footnote{The $E_2$ term in the Leray spectral sequence reads
$
  E_2^{p,q}\,=\,H^p(X) \otimes H^q( \mathbb{C}^* ) \, \simeq \left.\,
  \begin{cases} \mathbb{R}^{b_p} & q=0,1 \\ 0 & {\rm else} \end{cases} \right\} 
$ 
in terms of the Betti numbers $b_p = \dim H^p(X)$. The spectral sequence applied to $\mathbb{C}^*$-bundles degenerates at $E_3^{p,q}=H^{p+q}(\hat X)$, and the only non-trivial differential is $d_2: E_2^{p-2,1} \rightarrow E_2^{p,0}, \vartheta \mapsto (-1)^{p+1}\pi^*c_1(\mathbb{C}^*)\wedge \pi^*\vartheta$;  together, these imply that the two-form $\pi^*c_1(\mathbb{C}^*)=-d_2(\text{0-form on}\ \mathbb{C}^*)$ is exact in the space $\hat X$.}
 i.e.,
\be
   H^2(\hat X) \simeq H^2(X)/{c_1(\mathbb{C}^*)} \ .
\ee
This is in agreement with the observation that in the construction of the global one-form $\hat\theta$ in \eqref{eq:theta}, the $\mathbb{C}^*$-fiber directions play an essential role.

We can also check that the K\"ahler manifold $\hat X$ satisfies the general geometric criteria (collected at the end of Section \ref{sec:sigma-review}) for a K\"ahler manifold to admit an exact K\"ahler form. Namely, the K\"ahler manifold is non-compact since the $\mathbb{C}^*$-fibers are non-compact. Furthermore, since the $\mathbb{C}^*$-fibers are non-compact we can only try to construct compact holomorphic subspaces (of dimension greater than zero) in the base. However, any compact holomorphic subspace of the base $X$ ceases to be compact in the total space $\hat X$.  Owing to the fact that we require a positive-definite K\"ahler metric, the $\mathbb{C}^*$-fibration restricts to a non-trivial $\mathbb{C}^*$-fibration over any compact holomorphic subspace. Moreover, a non-trivial $\mathbb{C}^*$-fibration does not admit any global holomorphic sections. Therefore, the considered compact subspace of the base $X$ cannot be holomorphically embedded in the $\mathbb{C}^*$-fibered target space $\hat X$. 

We end this section with a discussion on the metric \eqref{eq:Khat-metric} of the target space $\hX$ and, in particular, on the structure of the real and sufficiently smooth function $H(\rho)$. As observed in \eqref{eq:ConH}, $H(\rho)$ must be a monotonically increasing and negatively curved function in order to yield a non-degenerate K\"ahler metric. The latter conditions on $H(\rho)$ are necessary to arrive at a \smodel\ with a well-defined, non-tachyonic kinetic term. Furthermore, because of the non-compactness of the target space $\hat X$ (due to the non-compact $\C^*$-fibers), we may choose the function $H(\rho)$ such that the target space boundary is at infinite distance. 

To arrive at the conditions for infinitely far boundaries, we first note that the domain of function $H(\rho)$ is a real interval $(\rho_-,\rho_+)$. The boundary of this domain corresponds to the target space boundary in the fiber direction.\footnote{Note that $\rho_-$ and $\rho_+$ could also be $-\infty$ or $+\infty$, respectively. Actually, in order to obtain a smooth function with $H'(\rho)>0$ and $H''(\rho)<0$ with a boundary at infinite distance, it turns out that one boundary of the domain of $H(\rho)$ must be infinite.}  To determine the distance to the boundary, it is convenient to rewrite the metric  \eqref{eq:Khat-metric} in terms of the variables $\rho$ and $a=\Im(y)$,
\be
ds_{\hX}^2 = -\frac{H''(\rho)}{4} d\rho^2 - H''(\rho) (da - \Im(k_i d\csc^i))^2 + H'(\rho) g_{i\jb} d\csc^id\bcsc^\jb \ .
\ee
Then the conditions for the boundary to be at infinity read
\be
  \int_{\rho_-}^{\rho_0} ds_{\hat X} = - \frac14 \int_{\rho_-}^{\rho_0} d\rho \, \sqrt{H''(\rho)} = \infty \ , \qquad
  \int_{\rho_0}^{\rho_+} ds_{\hat X} = - \frac14 \int_{\rho_0}^{\rho_+} d\rho \, \sqrt{H''(\rho)} = \infty \ ,
\ee
where $\rho_0$ is an arbitrary point in the interval $(\rho_-,\rho_+)$. These constraints, together with \eqref{eq:ConH}, are met by the function $H(\rho) = \log \rho$, for example, with the domain $(0,+\infty)$. This choice actually appears in the context of large volume compactifications in various string scenarios, as we discussed in section~\ref{sec:ModStab}.

\newpage

\providecommand{\href}[2]{#2}\begingroup\raggedright\endgroup

\end{document}